\documentclass[
    aps,
    prx,
    reprint,
    superscriptaddress,
    longbibliography,
    amsmath,
    amssymb,
    floatfix,
    nofootinbib
]{revtex4-2}
\usepackage[utf8]{inputenc}
\usepackage{amsfonts}
\usepackage{physics}
\usepackage{graphicx}
\usepackage[dvipsnames]{xcolor}
\usepackage[]{hyperref}
\usepackage{bm}
\usepackage[caption=false]{subfig}

\usepackage[T1]{fontenc}
\usepackage{float}
\usepackage{placeins}

\usepackage{tikz-network}
\usepackage{pgfplots}
\pgfplotsset{compat=newest}
\usepackage{tikz-3dplot}
\usetikzlibrary{fadings}

\usepackage{orcidlink}
\newcommand{\orcidsimone}{\orcidlink{0000-0002-8882-2169}}
\newcommand{\orcidgiuseppe}{\orcidlink{0000-0002-7280-445X}}
\newcommand{\orcidjulian}{\orcidlink{0000-0002-7011-6477}}
\newcommand{\orcidfrancesco}{\orcidlink{0000-0003-2008-5956}}
\newcommand{\orcidalberto}{\orcidlink{0000-0001-9112-8664}}
\newcommand{\orcidmarco}{\orcidlink{0000-0002-3215-3453}}
\newcommand{\orcidivano}{\orcidlink{0000-0001-5690-1981}}

\newcommand{\unipd}{Dipartimento di Fisica e Astronomia "G. Galilei", via Marzolo 8, Italy I-35131, Padova, Italy}
\newcommand{\padcen}{Padua Quantum Technologies Research Center, Universit{\`a} degli Studi di Padova}
\newcommand{\pdinfn}{Istituto Nazionale di Fisica Nucleare (INFN), Sezione di Padova, I-35131, Padova, Italy}
\newcommand{\ibmquantum}{IBM Quantum, IBM Research Europe - Zurich, S\"aumerstrasse 4, 8803 R\"uschlikon, Switzerland}
\newcommand{\epfl}{Institute of Physics, \'Ecole Polytechnique F\'ed\'erale de Lausanne (EPFL), CH-1015 Lausanne, Switzerland}
\newcommand{\unibari}{Dipartimento di Fisica, Universit\`a di Bari, I-70126 Bari, Italy}
\newcommand{\bariinfn}{Istituto Nazionale di Fisica Nucleare (INFN), Sezione di Bari, I-70125 Bari, Italy}

\begin{document}

\title{Hybrid tree tensor networks for quantum simulation}

\author{Julian Schuhmacher\orcidjulian}
\altaffiliation{These authors have contributed equally}
\affiliation{\ibmquantum}
\affiliation{\epfl}

\author{Marco Ballarin\orcidmarco}
\altaffiliation{These authors have contributed equally}
\affiliation{\unipd}
\affiliation{\padcen}
\affiliation{\pdinfn}

\author{Alberto Baiardi\orcidalberto}
\affiliation{\ibmquantum}

\author{Giuseppe~Magnifico\orcidgiuseppe}
\affiliation{\unipd}
\affiliation{\unibari}
\affiliation{\bariinfn}

\author{Francesco Tacchino\orcidfrancesco}
\affiliation{\ibmquantum}

\author{Simone Montangero\orcidsimone}
\affiliation{\unipd}
\affiliation{\padcen}
\affiliation{\pdinfn}

\author{Ivano Tavernelli\orcidivano}
\email[Electronic address: ]{ita@zurich.ibm.com}
\affiliation{\ibmquantum}

\begin{abstract}
    Hybrid Tensor Networks (hTN) offer a promising solution for encoding variational quantum states beyond the capabilities of efficient classical methods or noisy quantum computers alone.
    However, their practical usefulness and many operational aspects of hTN-based algorithms, like the optimization of hTNs, the generalization of standard contraction rules to an hybrid setting, and the design of application-oriented architectures have not been thoroughly investigated yet.
    In this work, we introduce a novel algorithm to perform ground state optimizations with hybrid Tree Tensor Networks (hTTNs), discussing its advantages and roadblocks, and identifying a set of promising applications.
    We benchmark our approach on two paradigmatic models, namely the Ising model at the critical point and the Toric code Hamiltonian.
    In both cases, we successfully demonstrate that hTTNs can improve upon classical equivalents with equal bond dimension in the classical part.
\end{abstract}

\maketitle

\section{Introduction}

The study of quantum many-body (MB) systems lies at the heart of contemporary physics, encompassing a broad range of application domains -- from condensed matter~\cite{Bruus2004_ManyBodyBook,Giuliani2005_QuantumTheoryElectronLiquid} to lattice gauge theories~\cite{Dalmonte2016_LatticeGauge,Cirac2017_DigitalLatticeGauge,banuls2020review,dimeglio2023quantum,PRXQuantum.4.027001} -- and posing some of the hardest challenges for computational simulations.
Prominent examples of state-of-the-art numerical methods employed to simulate MB systems are based on Tensor Network (TN) states~\cite{white1992density,Rommer1997,verstraete2004renormalization,shi2006classical,perez2006matrix,McCulloch2007,vidal2008class,evenbly2014class,Baiardi2020_DMRG-Review}, Neural Network Quantum States~\cite{carleo2017solving,Hermann2023_NNTS-Chemistry}, and Quantum Monte Carlo~\cite{ceperley1986,Sandvik1991,Troyer2005_FermionicMonteCarlo,Booth2009_FCIQMC}.
All of these techniques exploit some inherent structure exhibited by the target physical quantum states, e.g., the area law of the entanglement entropy for low-energy eigenstates of one-dimensional gapped Hamiltonians~\cite{hastings2007iop,eisert2010}.
However, when considering more complex problems, e.g. optimizing the ground state of Hamiltonians in 2 or more dimensions~\cite{Orus2019_Review} or encoding quantum states generated by long-time dynamical simulations~\cite{Zwolak2020_EntanglementBarrier}, most classical approaches eventually struggle.

In recent years, quantum computers have emerged as an alternative, promising platform for the study of quantum MB systems~\cite{ebadi2021quantum,scholl2021quantum,Miessen2022_QuantumDynamics-Review,Kim2023_QuantumUtility,shtanko2023uncovering}.
Both, digital quantum processors~\cite{tacchino2020quantum} and specialised analog platforms~\cite{daley2022practical} can in principle be used as quantum simulators to prepare strongly correlated quantum states and to simulate their time evolution.
However, despite remarkable progress, these devices are still limited in practice.

A natural and increasingly popular approach to alleviate this issue is to combine classical and quantum computations
~\cite{mcclean2016,slattery2021,araz2022,callison2022,eddins2022doubling,Schoulepnikoff2023,rudolph2023nature,martin2023,martin2023_BP,filippov2023scalable}.
By integrating state-of-the-art classical methods and constrained quantum resources, one may, for instance, partition the computational load such that the quantum hardware is only used for the most challenging simulation steps, e.g., to describe strong correlations~\cite{rossmannek2021quantum,rossmannek2023quantum}, while the rest of the simulation is offloaded to a classical computer.

\begin{figure}[t]
    \centering
    \includegraphics[width=0.95\columnwidth]{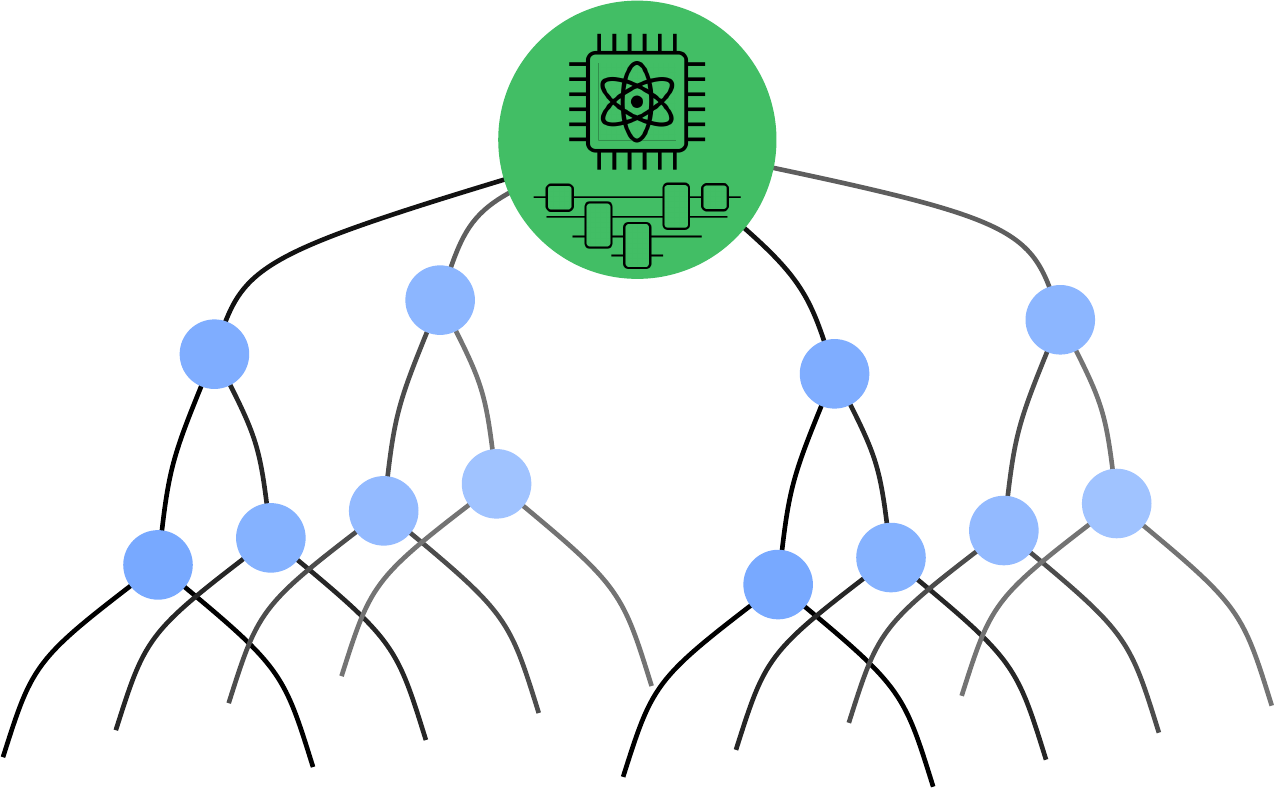}
    \caption{
        Hybrid Tree Tensor Network, where the classical tensors are identified by the blue color, and the top quantum tensor, colored in green, is pictorially represented by a quantum processor.
        }
    \label{fig:graphical_abstract}
\end{figure}

Along these lines, a particularly promising framework is offered by hybrid Tensor Networks (hTNs)~\cite{yuan2021quantum}, in which quantum states prepared on quantum processors are interpreted as quantum tensors and embedded -- with a set of suitable contraction rules -- within a classical TN.
Hybrid methods such as sub-space expansion~\cite{Takeshita2020_QSE}, entanglement-forging~\cite{eddins2022doubling} or deep-VQE~\cite{fujii2022deep} can be also considered as special instances of hTNs.
While the general theoretical foundations of hTNs have been set~\cite{yuan2021quantum,kanno2021quantum} -- including some preliminary studies on ground state optimization~\cite{yuan2021quantum,kanno2024quantum} and noise propagation~\cite{harada2023noise} -- many crucial aspects related to the practical use of hTNs remain to be investigated.
Notably, many routine procedures employed for classical TNs, such as isometrization and contraction, require substantial efforts to be ported in to a hybrid quantum-classical settings, and bear hidden costs that have up to now been neglected. 
Other open questions remain, such as identifying best practices for the optimization of hTNs, quantifying the overall scaling of different flavours of hTNs with system size, and characterizing the most promising applications (which may include quantum dynamics, quantum chemistry, and open quantum systems).

In this work, we identify a specific hierarchically structured hybrid ansatz, inspired by classical Tree Tensor Networks (TTNs) and pictorially described in Fig.~\ref{fig:graphical_abstract}, which allows for an effective integration of quantum tensors within classical tensor networks.
We then describe a scalable sweep-based algorithm for the optimization of general loop-less hTNs that mimics the corresponding routine employed for classical TTNs~\cite{gerster2014unconstrained}.
We resolve all theoretical hurdles arising from the translation of this procedure into the hTN framework, hence establishing a complete workflow for preparing and optimizing hTNs.
Finally, we showcase our proposed pipeline by computing the ground state of the transverse field Ising model and of the Toric code Hamiltonian, demonstrating numerically that hTNs can successfully enhance the accuracy of a corresponding, fully classical TN simulation.

A recently published preprint~\cite{huang2023tensor} also develops a TNs-based quantum algorithm. However, our work differs from theirs in the following aspects: (i) The setup considered in~\cite{huang2023tensor} consists of an unique quantum tensor, to which a unitary MPO is applied. Therefore, the resulting hTN contains loops. Conversely, we solely consider loop-less hTN, which can be optimized more efficiently. (ii) The classical tensors in the setup proposed in Ref.~\cite{huang2023tensor} are parameterized through quantum gates, whereas in our approach, no fixed parameterization is used to encode the classical tensors, as one would do for entirely classical TNs. (iii) In Ref.~\citenum{huang2023tensor}, the parameters of the quantum and classical tensors are optimized simultaneously, with global optimization algorithms. Our approach allows a local, iterative optimization of the tensors. It is known from classical TNs, that such a strategy is more scalable than a global one.

This manuscript is structured as follows.
In section~\ref{sec:methods}, we provide an introduction to TNs and hTNs and motivate the studied hTTN architecture.
In section~\ref{sec:optimization_algorithm}, we introduce a novel optimization algorithm for loop-less hTNs.
Simulations of the hTN optimization are presented in section~\ref{sec:results}.
The manuscript then concludes in section~\ref{sec:conclusion} with a summary and an outlook on future research directions.

\section{Methods}\label{sec:methods}

\begin{figure}
    \centering
    \subfloat[Rank-$n$ tensor\label{sfig:classical_tensor}]{
        \centering
        \includegraphics[width=0.75\columnwidth]{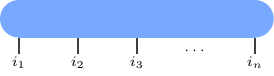}
    }\\
    \subfloat[MPS decomposition\label{sfig:classical_mps}]{
        \centering
        \includegraphics[width=0.75\columnwidth]{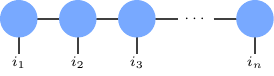}
    }
    \caption{Decomposition of a rank-$n$ tensor representing a general quantum state (a) as an MPS (b).}
    \label{fig:notation}
\end{figure}

The wave function describing the state of a $n$-body (referred to as sites in the following) quantum MB system with local dimension $d$ can be written as
\begin{equation}
  \ket{\psi} = \sum_{i_1,i_2, \ldots, i_n} \psi_{i_1 i_2 \ldots i_n} \ket{i_1} \ket{i_2} \cdots \ket{i_n} \,, 
  \label{eq:mb_wave_function}
\end{equation}
where $i_l \in \{1, 2 \ldots, d\}$, and $\ket{i_l}$ represents a set of (computational) basis states.
The state $\ket{\psi}$ is an element of the Hilbert space $\mathcal{H} = (\mathbb{C}^{d})^{\otimes n}$ with total dimension $d^n$. 
Hence, storing the amplitudes $\psi_{i_1 i_2 \ldots i_n}$ is, in general, exponentially inefficient for increasing $n$.
Techniques such as the Density Matrix Renormalization Group (DMRG)~\cite{white1992density,white1993density,schollwock2005density} and TNs~\cite{Rommer1999,verstraete2004renormalization,perez2006matrix,shi2006classical,McCulloch2007,vidal2008class,Rizzi2008,Cincio2008,Tagliacozzo2009,Murg2010,evenbly2014class,orus2014practical,Schollwoeck2011,montangero2018introduction,silvi2019tensor} address this problem by compressing $\psi_{i_1 i_2 \ldots i_n}$ into a collection of low-rank tensors.
In the graphical notation used throughout this article, a rank-$n$ tensor such as $\psi_{i_1 i_2 \ldots i_n}$ is visualized as a shape with $n$ legs attached to it (see Fig.~\ref{sfig:classical_tensor}), where each leg corresponds to a single index.
Similarly, TNs are represented by multiple shapes connected together.
For example, a Matrix Product State (MPS)~\cite{Rommer1999,perez2006matrix,McCulloch2007,Schollwoeck2011} is represented in Fig.~\ref{sfig:classical_mps} as a set of rank-3 tensors arranged along a line. 
As shown in Fig.~\ref{sfig:classical_mps}, the tensors composing a TN have two types of legs: \textit{physical} ones that correspond to the wave function indices $(i_1, i_2, \ldots, i_n)$ and are represented with downward facing segments in Fig.~\ref{fig:notation}, and \textit{internal} links that connect different tensors in the TN and symbolize their contraction. 
The local dimension associated to internal links is called bond dimension and is denoted as $\chi$. 
Importantly, a TN-based description of a quantum state is considered effective if a reasonably small $\chi$ is sufficient to describe the state.
For example, while $\chi = d^{\lfloor n/2 \rfloor}$ is required to exactly describe an arbitrary wave function in the form of an MPS, one-dimensional systems respecting the area law~\cite{verstraete2006matrix,hastings2007entropy,eisert2010,arad2013area} only require a fixed maximum bond dimension $\chi$ for increasing system size to be represented with a given accuracy.

\subsection{Hybrid tensor networks}

\begin{figure}
    \centering
    \subfloat[\label{sfig:quantum_tensor}]{
        \centering
        \includegraphics[width=0.55\columnwidth]{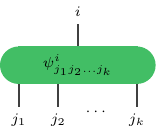}
    }
    \hfill
    \subfloat[\label{sfig:quantum_tensor_contraction}]{
        \centering
        \includegraphics[width=0.35\columnwidth]{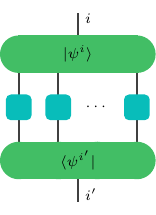}
    }
    \caption{
    (a) Diagrammatic representation of a quantum tensor of rank $(k+1)$. 
    Quantum tensors are highlighted in green.
    (b) Contraction of the quantum tensor in (a) with an observable $O$.}
    \label{fig:quantum_tenor_diagrams}
\end{figure}

Storing and manipulating a tensor on classical hardware becomes inefficient for increasing bond dimension. 
To overcome this limitation, in the hTNs framework~\cite{yuan2021quantum} high-rank tensors are represented by quantum states prepared on quantum computers.
A rank-$(k+1)$ quantum tensor $\psi^i_{j_1 j_2 \ldots j_k}$ is implemented as
\begin{equation}
    | \psi^i \rangle = | \psi^i_{j_1 j_2 \ldots j_k} \rangle = \sum_{j_1 j_2 \ldots j_k} \psi^i_{j_1 j_2 \ldots j_k} \ket{j_1} \ket{j_2} \cdots \ket{j_k} \,,
    \label{eq:quantum_tensor}
\end{equation}
where $j_l$ are the quantum indices corresponding to qubit registers, and $i$ is a classical index labeling the set of independent quantum states $\{\ket{\psi^i}\}$.
In principle, a quantum tensor may have multiple classical indices.
However, we will limit the following derivation to quantum tensors with a single classical index, because all quantum tensor operations can straightforwardly be applied to multiple classical indices, once they are combined into a single index.
Note that, we have also introduced the convention to write the quantum indices of a tensor as a subscript, and the classical ones as a superscript.
In the diagrammatic notation introduced in the previous section, we will denote quantum tensors in green and classical tensors in blue.
Fig.~\ref{sfig:quantum_tensor} shows the quantum tensor given in Eq.~\eqref{eq:quantum_tensor}, where the $k$ quantum indices $j_l$ are pointing downwards, and the classical index $i$ is pointing upwards.

Throughout this work, we use parameterized quantum circuits of the form $U^i(\bm{ \theta}_i)$ to prepare the quantum states representing a quantum tensor
\begin{equation}
    | \psi^i \rangle = U^i(\bm{\theta}_i) \ket{0} \,,
\end{equation}
where the circuit unitary $U^i$ and the set of parameters $\bm{\theta}_i$ change for different values of the classical index $i$s (see Refs.~\cite{yuan2021quantum,harada2023noise} for other methods to represent quantum tensors).

The potential of quantum tensors only becomes clear once we introduce appropriate contraction rules, as originally proposed in Ref.~\cite{yuan2021quantum}, to combine multiple quantum tensors and connect them to their classical counterparts. 
Two classes of contraction rules for quantum tensors exist, depending on whether the contracted index is a classical or a quantum one.
Contracting a quantum tensor $|\psi^i \rangle$ with a classical tensor $T^{i i'}$ over the index $i$ through the operation
\begin{equation}
    | \tilde{\psi}^{i'} \rangle = \sum_i T^{i i'} | \psi^i \rangle \, ,
\end{equation}
resulting in a set of states that are linear combinations of the original states $\{\ket{\psi^i}\}$. 
Two quantum tensors $|\psi^i \rangle$ and $| \phi^i \rangle$ may also be contracted over a classical index by taking a linear combination of the tensor products
\begin{equation}
    | \tilde{\psi} \rangle = \sum_i | \psi^i \rangle \otimes | \phi^i \rangle \,.
\end{equation}
If, instead, the contracted index of the quantum tensor $| \psi_{j_1 j_2 \ldots j_k} \rangle$ is a quantum one $j_l$, we first convert it into a classical index, $j_l \rightarrow i$, through a projection, defining the set of un-normalized states $\{| \psi^i \rangle = \langle i | \psi \rangle\}$.
These correspond to a rank-$k$ quantum tensor ($(k-1)$ quantum indices and one classical index) which can be contracted with another tensor using the rules defined above.

Besides contractions, the most important operation that can be performed with a quantum tensor is calculating expectation values.
For an individual quantum state $\ket{\psi}$, i.e.\ a quantum tensor with only quantum indices, the expectation value of an observable $O = O_1 \otimes O_2 \otimes \cdots \otimes O_k$ is defined as 
\begin{equation}
  \langle O \rangle = \bra{\psi} O_1 \otimes O_2 \otimes \cdots \otimes O_k \ket{\psi} \,,
  \label{eq:ExpectationValue_Q}
\end{equation}
where each $O_i$ is an Hermitian operator with support on the quantum register associated with the $i$-th quantum leg.
However, when a quantum tensor is embedded within an hTN, some of its classical or quantum legs may not be involved in the contraction with $O$ and are, instead, either left open (i.e., uncontracted) or contracted with other tensors.
For example, in Fig.~\ref{sfig:quantum_tensor_contraction}) we depict the case in which two links ($i$ and $i'$ in the figure) are left uncontracted in the calculation of an expectation value.
If the open indices are classical, the contraction is evaluated as
\begin{equation}
  M^{i' i} = \langle \psi^{i'} | O_1 \otimes O_2 \otimes \cdots \otimes O_k | \psi^i \rangle \,.
  \label{eq:measurement_matrix_classical}
\end{equation}
If, instead, the open indices are quantum, the contracting yields the following expectation value
\begin{equation}
    M^{i' i} = \bra{\psi} (\ketbra{i'}{i}) \otimes O_1 \otimes O_2 \otimes \cdots \otimes O_k \ket{\psi} \,.
     \label{eq:measurement_matrix_quantum}
\end{equation}
Note that the projector $\ketbra{i'}{i}$ transforms the quantum indices to classical ones.
Therefore, in both cases above, the outcome is a classical matrix $M$, which can then be further contracted with other tensors.

The entries of the matrix $M$ are evaluated by executing the quantum circuits given in Ref.~\cite{yuan2021quantum}.
Note that Ref.~\cite{yuan2021quantum} describes only the procedure to contract open quantum indices of bond dimension $\chi=2$, which can be implemented with a single qubit.
In appendix~\ref{app:open_link_contraction}, we generalize this procedure to a generic bond dimension of $\chi = 2^k$, implemented with $k$ qubits.

\medskip

At this point, it is important to highlight that, even though the hTN formalism described above allows one to implement arbitrary contractions between classical and quantum tensors, these operations are not necessarily efficient if executed explicitly. 
Indeed, in many cases one would need access to all entries of a tensor -- a task that, for quantum components, is akin to full tomography.
Conversely, calculating expectation values on quantum tensors essentially replaces some of these explicit contractions with a renormalization of $O$ followed by sampling.
The latter is efficiently implementable under reasonable sparsity assumptions on the (renormalized) observable.
This implies that the order in which an hTNs is contracted is crucial: in general, the most efficient contraction is obtained by first contracting all classical indices, by leaving the contractions over the quantum ones implicit until they are evaluated through an expectation value calculation.
A detailed investigation on the effect of the contraction order on the algorithm efficiency is presented in Appendix~\ref{app:contraction_order}.

\subsection{Hybrid Tree Tensor Networks}

One of the aims of this work is to pinpoint instances where hTNs outperform their classical counterparts.
A natural setup to investigate them is provided by scenarios where a quantum encoding of tensors is employed as soon as their classical description becomes unfeasible.
In the following, we consider such a setup for the hierarchically structured ansatz family of Tree Tensor Networks (TTNs)~\cite{shi2006classical,Tagliacozzo2009,Murg2010}, which we will now quickly revisit and later promote to their hybrid version.

A TTN is a loop-less TN, i.e., one in which each tensor is connected to any other tensor with a unique path.
TTNs display a hierarchical structure inspired by the renormalization group~\cite{shi2006classical}, and are effective for the description of physically relevant states in one-dimensional systems that feature an entanglement entropy following the area-law~\cite{hastings2007entropy,arad2013area}.
Compared to MPSs, they generally require a lower $\chi$ to represent the same quantum state, especially for $2$- or higher-dimensional systems.
However, TTNs do not satisfy the area law in 2 dimensions, as opposed to other approaches, e.g. projected entangled pair states (PEPSs)~\cite{verstraete2004renormalization}.
Nevertheless, loop-less TNs yield several advantages over PEPSs, e.g., the possibility to efficiently compute expectation values or the availability of optimization algorithms with a more favourable scaling in the bond dimension~\cite{orus2014practical, Czarnik2015, lubash2014, Cataldi2021hilbertcurvevs}).
For this reason, several techniques have been developed to improve the effectiveness of TTNs as, e.g., augmented or adaptive TTNs~\cite{felser2021efficient, quian2022, PhysRevB.105.214201} allowing simulations of high-dimensional systems of moderate sizes.

\begin{figure}
    \centering
    \subfloat[\label{sfig:classical_ttn}]{
        \centering
        \includegraphics[width=0.30\columnwidth]{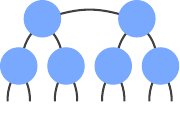}
    }%
    \subfloat[\label{sfig:hybrid_ttn}]{
        \centering
        \includegraphics[width=0.30\columnwidth]{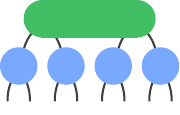}
    }%
    \subfloat[\label{sfig:multiple_qtensors_ttn}]{
        \centering
        \includegraphics[width=0.30\columnwidth]{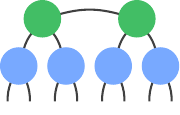}
    }
    \caption{
        Two-layer classical (a) and hybrid (b-c) TTN.
        Compared to the classical TTN, the hybrid TTN in (b) is constructed by replacing the topmost layer with a single quantum tensor, while the one in (c) by replacing each classical tensor at the top with a quantum tensor.}
    \label{fig:ttns}
\end{figure}

In the following, we focus on \textit{binary} TTNs, even though all our findings can straightforwardly be generalized to an arbitrary TTN structure.
A binary TTNs (see Fig.~\ref{sfig:classical_ttn}) is composed of rank-3 tensors that are organized in layers.
For a tensor in a given layer, two legs are connected to tensors in the layer below, and one leg is connected to a single tensor in the layer above.
The only exceptions appear in the lowest layer, where two legs per tensor representing the physical links are left uncontracted, and in the topmost layer, where the two tensors are connected to each other.
As we traverse a TTN from bottom to top, the individual tensors support and describe quantum correlations between increasingly more sites.
This generally increases the bond dimension necessary to accurately encode the quantum state.
The bond dimension required to exactly (i.e., without any truncation) represent a general quantum state grows from $\chi$ to $\chi^2$ when going from one layer to the next~\cite{Schollwoeck2011}.
Thus, the maximal bond dimension required to exactly represent a general state on $n$ sites with local dimension $d$ is $\chi_{max}=d^{2n_l}$, where $n_l=\log_2 n-1$ is the number of layers of the TTN.
Whereas for one-dimensional systems a much more favourable scaling is guaranteed by the area law, this worst-case scaling may be encountered in higher dimensions.
In fact, for Hamiltonians in 2 or more dimensions, the bond dimension required to accurately capture the important correlations in large-size systems can easily exceed what is currently achievable classically~\cite{Williams2020_ManyBodyBenchmark,Eriksen2020_Benzene}.
It is therefore reasonable to ask whether placing one or more quantum tensors at the top of a classical TTN could overcome the current limits of classical TNS-based quantum simulation.

Our proposed hybrid TTN (hTTN) architecture is schematically illustrated in Fig.~\ref{sfig:hybrid_ttn} (note that a similar structure was briefly mentioned, but not thoroughly investigated, in Ref.~\cite{yuan2021quantum}).
We define the \textit{virtual} bond dimension $\chi_{\text{virt}}$ of the quantum tensor at the top as the bond dimension required to represent it exactly with a classical TN. 
A quantum tensor defined on 40 qubits can, at least in principle, realize virtual bond dimensions of $\chi_{\text{virt}} \sim 2^{20}$, which is far beyond the highest bond dimensions reported in recent classical TN literature~\cite{Brabec2021_MassivelyParallel-DMRG,Liu2023,oh2023,Xiang2024_GPU-DMRG}.

hTNs with different topologies have already been studied.
For example, an architecture in which two quantum subsystems are described with quantum tensors and their (weak) correlation is encoded via a classical tensor -- a procedure also known as \textit{entanglement forging} -- was implemented in Refs.~\cite{eddins2022doubling,huembeli2022entanglement,harada2023noise}.
However, unlike our proposed hTTNs, this approach is tailored to quantum systems where long-range quantum correlations are weak, and can be efficiently approximated classically.
Another option are hTNs that consists of quantum tensors only~\cite{yuan2021quantum}.
Compared to these schemes, we believe that our proposed setup offers the most effective and generally applicable design, primarily because the bond dimension will generally grow towards the top of a TTN.
This in-principle advantage can only become concrete -- and its potential limitations can only be correctly characterized -- if the abstract hTTN structure is equipped with scalable optimization procedures, and if the bottlenecks arising at quantum-classical interfaces are properly accounted for, as introduced hereafter.

\section{Optimization of loop-less \MakeLowercase{h}TN\MakeLowercase{s}}
\label{sec:optimization_algorithm}

In Ref.~\cite{yuan2021quantum}, the authors make use of a global optimization strategy for hTNs based on quantum imaginary time-evolution.
The general applicability of this approach is limited by two main factors.
First, it only supports hTNs consisting solely of quantum tensors.
Second, the simultaneous optimization of all parameters in large variational quantum circuits is known to suffer from trainability issues~\cite{thanasilp2023subtleties} and it does not correspond to state-of-the-art classical strategies, relying on local, iterative optimization methods.

To address these points, we introduce an alternative strategy based on the DMRG-inspired optimization algorithm for classical TTNs~\cite{gerster2014unconstrained}.
The underlying idea, borrowed from the classical routine, is to replace the global optimization of the full (h)TTN with a sequential, iterative optimization of each individual tensor.
This greatly reduces the computational complexity of the optimization, at the possible cost of a slower convergence.
Our novel procedure, adapted to hTNs, will be described in the subsequent sections.
To set the notation and the terminology, we will first briefly review the key steps of classical TTN optimization.

\begin{figure}[t!]
    \centering
    \includegraphics[width=\columnwidth]{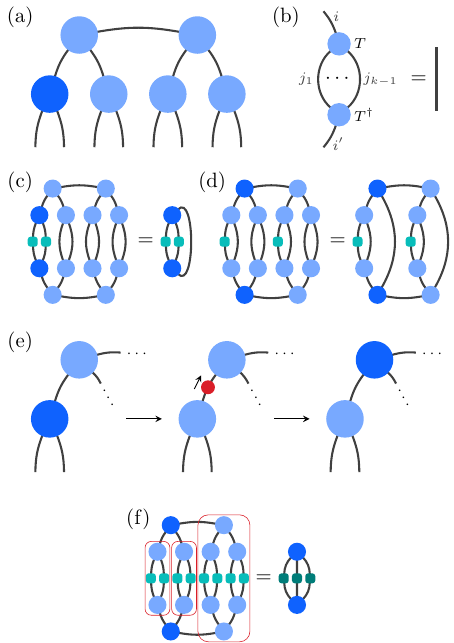}
    \caption{
        Classical TTN operations in diagrammatic notation. 
        (a) two-layer tree tensor network with the isometrization center at the leftmost tensor in the bottom layer (dark blue). 
        (b) Isometrization condition for all tensors except the isometrization center. 
        The indices are specified here, to show the correspondence with Eq.~\eqref{eq:isometry_condition}.
        In all other subfigures, the direction of the isometrization can be inferred from the location of the isometrization center.
        (c) Expectation value of a local two-site observable on the first two sites.
        The expectation value reduces to the contraction of a single tensor with the observable (teal). 
        (d) Expectation value of a general two-site observable on non-neighbouring sites.
        The isometrization center is placed at the highest tensor in the path that connects the two sites.
        The calculation reduces to a contraction of the tensors along this path. 
        (e) Shift of the isometrization center between the two tensors at the left edge of the TTN in (a).
        The red dot represents the non-isometric part of the bottom tensor (arising from the QR decomposition) that is contracted with the top tensor, making the top tensor the new isometrization center.
        (f) Local effective Hamiltonian for the isometrization center at the top left tensor, each obtained from the contraction of the tensors in the red boxes.
    }
    \label{fig:classical_ttn_optimization}
\end{figure}

A gauge freedom exist when representing a quantum state as a TN.
In fact, contracting an index of a given tensor by a generic unitary matrix $U$ and the corresponding neighbouring tensor by $U^\dagger$ modifies the TNS representation without changing the underlying wave function.
For TTNs, this gauge freedom can be exploited by choosing a specific tensor in the TTN as the \textit{center of isometrization} towards which all other tensors are isometrized~\cite{gerster2014unconstrained}.
Specifically, all tensors $T_{i,(j_1 \ldots j_{k-1})}$ (except the center one) fulfill the isometry condition
\begin{equation}
  \sum_{j_1 \ldots j_{k-1}} T^\dagger_{i',(j_1 \dots j_{k-1})} T_{i, (j_1 \ldots j_{k-1})} = (\mathbb{I})_{i' i}
  \label{eq:isometry_condition}
\end{equation}
with respect to the index $i$ that points towards the center tensor.
Since a TTN does not contain any loops, this direction is uniquely defined.
For example, in the TTN represented in Fig.~\ref{fig:classical_ttn_optimization}a, the bottom left tensor is the isometrization center (highlighted in dark blue).
This implies that all other tensors in the TTN fulfill the isometry condition in Eq.~\eqref{eq:isometry_condition}, as illustrated in Fig.~\ref{fig:classical_ttn_optimization}b.
This implies that the norm of the full TTN (which is 1 for normalized quantum states) corresponds to the Frobenius norm of the center tensor, since all other tensors contract to the identity with their complex conjugate.
Concretely, the tensor $T_{j_1 \ldots j_k}$ representing the isometrization center fulfills the \emph{normalization} condition
\begin{equation}
  \sum_{j_1 \ldots j_k} T^\dagger_{j_1 \ldots j_k} T_{j_1 \ldots j_k} = 1 \,.
  \label{eq:normalization_condition}
\end{equation}
For later reference, we stress that, unlike the isometry condition of Eq.~\eqref{eq:isometry_condition}, where one index is left uncontracted, the sum in the normalization conditions includes all the indices.
Unless the open indices are trivial (i.e. 1-dimensional), the two conditions are mutually exclusive, since summing Eq.~\eqref{eq:isometry_condition} over the indices $i',i$ yields the trace of the identity matrix and, therefore, the index dimension.

Properly choosing the isometrization center can greatly simplify the evaluation of expectation values of local observables.
For instance, calculating the expectation value of a two-local observable on the first two sites of the TTN in Fig.~\ref{fig:classical_ttn_optimization}c, reduces to a contraction that only involves the center tensor.
To calculate the expectation value of a general two-local observable acting on non-neighbouring sites, the isometrization center is placed at the highest tensor along the path that connects the two sites.
The contraction of the observable with the full TTN then reduces to a contraction that only involves the tensors along the path connecting the two sites (see, e.g., Fig.~\ref{fig:classical_ttn_optimization}d).
Crucially, the isometrization center can be shifted between two neighbouring tensors via a QR decomposition~\cite{gerster2014unconstrained,Schollwoeck2011} of the current center tensor, where $Q$ is an isometry and $R$ is a matrix.
The matrix $R$ is then contracted with the neighbouring tensor, turning it into the new isometrization center.
This process is illustrated in Fig.~\ref{fig:classical_ttn_optimization}e for the two tensors at the left edge of the TTN in Fig.~\ref{fig:classical_ttn_optimization}a.
To shift the isometrization center between two non-neighbouring tensors, the isometrization center is moved along the path that connects them.

The second key concept in the TTN optimization algorithm is the \textit{local effective Hamiltonian} $\hat{H}_{\text{eff}}$ associated with a given tensor, which is obtained by contracting each of the branches departing from the isometrization center with the respective part of the original Hamiltonian $\hat{H}$.
Figure~\ref{fig:classical_ttn_optimization}f shows how the local effective Hamiltonian is obtained if the isometrization center is at the top left of the TTN for an Hamiltonian $\hat{H}$ which (for demonstration purposes) has support on all sites.
The local effective Hamiltonian plays a key role for the optimization of the center tensor. 
Specifically, at every step of the optimization procedure, the optimal center tensor is the eigenvector $T$ corresponding to the lowest eigenvalue $\epsilon$ of $\hat{H}_{\text{eff}}$ (see Ref.~\cite{gerster2014unconstrained} for derivation),
\begin{equation}\label{eq:opt_classical_tensor}
    \hat{H}_{\text{eff}} T = \epsilon T \,.
\end{equation}

The ground-state TTN representation is calculated by sequentially optimizing each tensor in the tree by first shifting the isometrization center to it, recalculating the local effective Hamiltonian, and optimizing the tensor.
Note that the local effective Hamiltonian does not have to be recalculated from scratch after shifting the isometrization to a neighbouring tensor.
It is sufficient to recalculate a single contraction that only involves both the tensor and the local effective Hamiltonian of the previous isometrization center (see Ref.~\cite{gerster2014unconstrained} for details).
One optimization sweep of the TTN is completed after optimizing once each tensor.
To extend this protocol to hTTNs, the key operations introduced above must be translated into equivalent ones for quantum tensors, as discussed below.

\subsection{Shift of the isometrization center in hybrid Tree Tensor Networks}
\label{ssec:implicit_isometrization}

A rank-$k$ quantum tensor $\psi_{j_1 \ldots j_k}$ including solely quantum indices fulfills the normalization condition in Eq.~\eqref{eq:normalization_condition} by construction.
Any update of the quantum tensor (e.g. by varying the parameters) will, therefore, always yield a normalized tensor.
Since it is not possible for a tensor to respect the isometry and the normalization condition simultaneously, one cannot straightforwardly extend the QR decomposition based algorithm discussed above to quantum tensors.

We therefore propose the following procedure to implicitly isometrize a quantum tensor:

\begin{enumerate}
    \item Calculate the open-link contraction of the quantum tensor and its complex conjugate, where the open link $i$ is the leg with respect to which we isometrize,
    \begin{equation}
        \sum_{\bm{j}} \psi^\dagger_{i',\bm{j}} \psi_{i,\bm{j}} = M^{i'i} \,,
    \end{equation}
    where $\bm{j} = (j_1 \ldots j_{k-1})$ collects the $(k-1)$ indices that are contracted.
    This corresponds to Eq.~\eqref{eq:measurement_matrix_classical} or~\eqref{eq:measurement_matrix_quantum}, depending on whether $i$ is a classical or quantum index, respectively.
    \item $M$ is a positive semi-definite Hermitian matrix (see Appendix~\ref{app:positive_semi_definiteness}) and, therefore, can be diagonalized through a unitary matrix $U$ such that $M = U^\dagger D U$. 
    Here, $D$ is the diagonal matrix containing the non-negative, real eigenvalues of $M$.
    \item Define $R = \sqrt{D} U$ as the non-isometric part of the quantum tensor, and $Q^i_{\bm{j}} = \sum_l \psi_{l,\bm{j}} (R^{-1})^{li}$ as the isometry.
\end{enumerate}

We define this isometrization as ``implicit'' because the quantum tensor itself does not change, but its contraction with the classical matrix $R^{-1}$ effectively acts as an isometry.
Indeed, we can readily verify that

\begin{equation}
 \begin{aligned}
    \sum_{\bm{j}} (Q^\dagger)^{i'}_{\bm{j}} Q^i_{\bm{j}} &= \sum_{\bm{j}} \sum_{l' l} ((R^{-1})^\dagger)^{i'l'} \psi^\dagger_{l',\bm{j}} \psi_{l,\bm{j}} (R^{-1})^{li} \\
    &= \sum_{l' l} ((\sqrt{D})^{-1} U)^{i'l'} M^{l'l} (U^\dagger (\sqrt{D})^{-1})^{li} \\
    &= ((\sqrt{D})^{-1} D (\sqrt{D})^{-1})^{i'i} = (\mathbb{I})^{i'i} \,,
 \end{aligned}
\end{equation}
where we followed the index convention introduced in section~\ref{sec:methods}.
To avoid numerical instability, we calculate $(\sqrt{D})^{-1}$ in practice from the pseudo-inverse of $D$.

\begin{figure}
  \centering
  \includegraphics[width=\columnwidth]{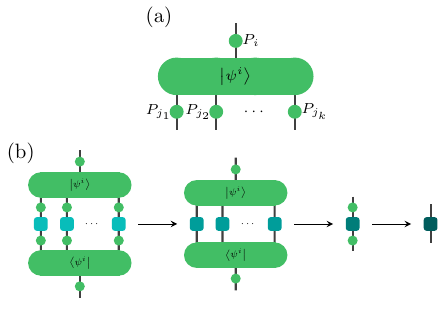}
  \caption{Diagrammatic representation of relevant quantum tensor operations.
  (a)~Extensions of a rank-($k$+1) quantum tensor with a set of classical matrices $\{P_l\}$ that are contracted with its legs.
  (b)~Contractions required to update the local effective Hamiltonian after isometrizing a quantum tensor.
  The set of classical matrices corresponding to the quantum tensor are either contracted with the observable (step 1$\rightarrow$2) or contracted with the results of the quantum tensor contraction (step 3$\rightarrow$4).}
  \label{fig:hybrid_ttn_optimization}
\end{figure}

In the rest of the paper, we combine all quantum tensors with a set of classical matrices $\{P_l\}_{l=1}^k$ that keep track of their (implicit) isometrization.
Each classical matrix is contracted with the relevant classical observables when calculating expectation values (see Sec.~\ref{ssec:local_eff_ham}).
The classical matrices associated to a given quantum tensor are also used to absorb the non-isometric parts originating when the isometrization center is moved from a neighbouring tensor to it.
In the diagrammatic notation, we symbolize the presence of a classical matrix through a circle connected to the legs of a quantum tensor to (see Fig.~\ref{fig:hybrid_ttn_optimization}a).
If the circle is absent, then $P_l = \mathbb{I}$.

Note that the presence of the classical matrices associated with a quantum tensor does not add significant computational overhead associated with the calculation of expectation values.
The tensor product structure of the classical matrices allows to straightforwardly integrate them into the contractions that involve a quantum tensor (see next section for an example).

\subsection{Definition of the local effective Hamiltonian for hybrid Tree Tensor Networks}
\label{ssec:local_eff_ham}

The local effective Hamiltonian can be calculated with quantum tensors as described above for classical tensors, calculating the contraction shown in the red boxes in Fig.~\ref{fig:classical_ttn_optimization}f.
For a quantum tensor, this corresponds to an open-link contraction performed according to Eq.~\eqref{eq:measurement_matrix_classical} or~\eqref{eq:measurement_matrix_quantum}, depending on whether the open link is classical or quantum, respectively.
Figure~\ref{fig:hybrid_ttn_optimization}b illustrates how to perform this contraction for a quantum tensor isometrized with respect to the topmost leg.
Note that the classical matrices $\{P_l\}$ introduced in the previous section are implicitly contracted with the quantum tensor either by a multiplying them with the target observable (step 1$\rightarrow$2) or with the operator obtained by contracting the quantum tensor with the observable (step 3$\rightarrow$4).

The considerations regarding the optimal order for contracting tensors and operators described above also hold when calculating the effective Hamiltonian.
For the open quantum-link contraction, we are in the unfavorable case in which a complete tomography of the link must be performed (see Appendix~\ref{app:contraction_order}).
Unfortunately, this step is unavoidable because the operator resulting from the tomography is the effective Hamiltonian required for the optimization of the neighbouring tensor. 
Since we only perform the tomography on a single leg at a time, the cost only scales with the number of qubits representing a leg, and not the number of qubits representing the full quantum tensor.

\subsection{Local optimization}
\label{ssec:local_opt}

We optimize each individual quantum tensor variationally. 
In the most general case, a rank-$k$ quantum tensor is encoded by a parameterized quantum state $\ket{\psi(\bm \theta)}$ and a set of classical matrices $\{P_l\}_{l=1}^k$ which we jointly write as $P = P_1 \otimes P_2 \otimes \cdots \otimes P_k$.
We define the following loss function
\begin{equation}
 \begin{aligned}
    \mathcal{L}(\bm \theta) &= \bra{\psi(\bm \theta)} P^\dagger \hat{H}_{\text{eff}} P \ket{\psi(\bm \theta)} \\
                        &+ \lambda \left| \bra{\psi(\bm \theta)} P^\dagger P \ket{\psi(\bm \theta)} - 1 \right|
 \end{aligned}
 \label{eq:loss_with_penalty}
\end{equation}
based on which we optimize the parameters $\bm \theta$. 
The first term is the expectation value of $\hat{H}_{\text{eff}}$, while the second term acts as a penalty term and is added to ensure the normalization of the quantum tensor (including classical matrices $\{P_l\}$).
If all classical matrices $\{P_l\}$ are unitary, the penalty term is trivially zero, and the loss function reduces to 
\begin{equation}
   \mathcal{L}(\bm \theta)
     = \bra{\psi(\bm \theta)} P^\dagger \hat{H}_{\text{eff}} P \ket{\psi(\bm \theta)} \,.
  \label{eq:loss_without_penalty}
\end{equation}
However, since the classical matrices are derived from the non-isometric part of the neighbouring tensors, they are generally non-unitary.

In both, equation~\eqref{eq:loss_with_penalty} and~\eqref{eq:loss_without_penalty}, the presence of the classical matrices $P$ associated with the legs of the quantum tensor transform every operator $\hat{O}$, such as a component of the effective Hamiltonian, into observables spanning all $k$ legs of the quantum tensor.
The reason for this is that the $\{P_l\}$ matrices are generally non-unitary, and therefore lead to a contribution to the observable for the quantum tensor even if the underlying observable only has support on a few legs.
In the worst case, the decomposition of a transformed observable as a sum of Pauli strings spans the whole $\chi^k \times \chi^k$ Hilbert space.
However, in Appendix~\ref{app:local_diag_of_obs} we present a strategy that exploits the tensor product structure of the $P$ matrices for reducing this steep scaling to a linear scaling in $k$ when calculating general expectation values.
Additionally, we propose a strategy in section~\ref{ssec:opt_workflow} which completely removes this overhead during the optimization of the quantum tensor by projecting each $P_l$ to its closest unitary.

\subsection{Definition of the optimization workflow}
\label{ssec:opt_workflow}

\begin{figure*}
    \centering
    \includegraphics[width=0.85\textwidth]{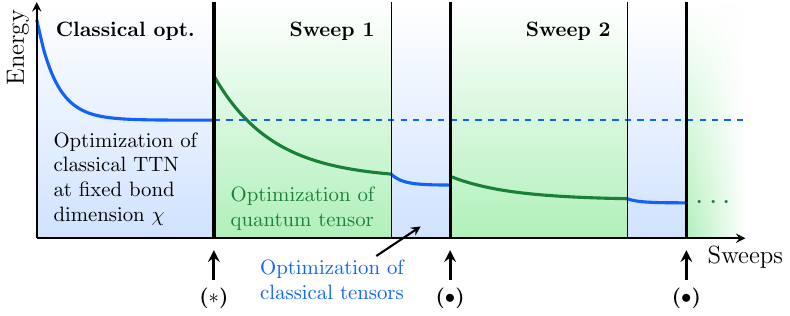}
    \caption{
        Proposed workflow for the optimization of hTNs.
        The workflow is sketched for a hTTN where the top layer is a single quantum tensor and all other layers are composed of classical tensors (see Fig.~\ref{sfig:hybrid_ttn}).
        The first step is the optimization of the corresponding classical TTN at a fixed, maximal bond dimension $\chi$.
        At \textbf{($\ast$)}, the quantum tensors in the hTTN are initialized by approximately mapping the tensors in the optimized TTN to a quantum circuit.
        The approximation may induce a discontinuity in the optimization profile.
        For each optimization sweep of the hTTN, we start by optimizing the quantum tensor at the top (green region), followed by the optimization of all classical tensors (blue region).
        At \textbf{($\bullet$)}, different strategies, also with different degrees of approximation, can be followed for preparing the quantum tensor for its next optimization (see section~\ref{ssec:opt_workflow} for details).
        The workflow can straightforwardly be generalized to different hTN structures, including architectures with multiple quantum tensors.
    }
    \label{fig:optimization}
\end{figure*}

\begin{figure}
    \centering
    \subfloat[Ladder quantum circuit\label{sfig:ladder_circuit}]{
        \centering
        \includegraphics[width=0.9\columnwidth]{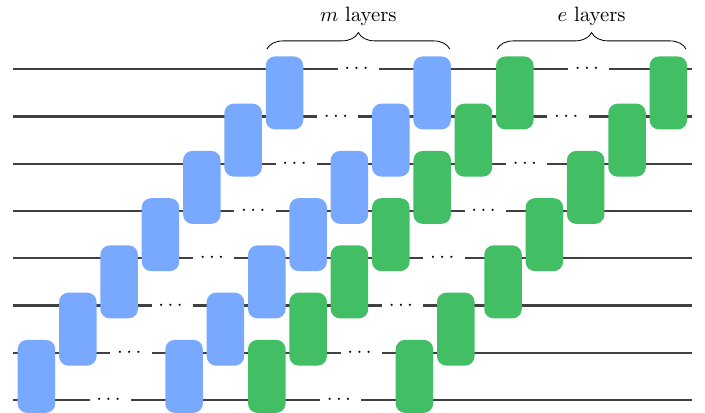}
    }\\
    \subfloat[Brick-wall quantum circuit\label{sfig:brick_wall_circuit}]{
        \centering
        \includegraphics[width=0.62307\columnwidth]{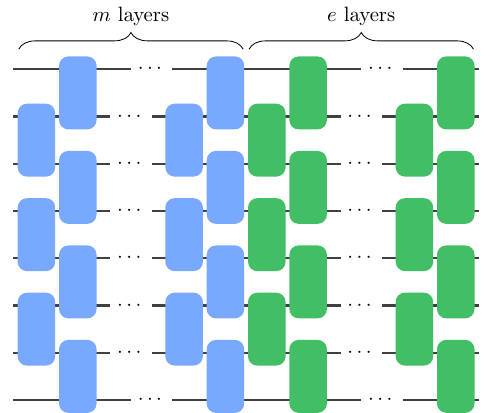}
    }
    \caption{Quantum circuit topologies applied to represent the quantum tensors.
    Both the ladder (a) and brick-wall (b) quantum circuits are built from $m$ layers initialized using the methods in Refs.~\cite{ran2020encoding,rudolph2023decomposition}, and $e$ layers used to extend the circuits.
    The gates in the $e$ additional layers are initialized close to identities.}
    \label{fig:quantum_circuits}
\end{figure}

Through the operations described in the previous sections, the classical TTN optimization algorithm can be extended to hTTNs.
Specifically, for a hTTN where the top layer is replaced with a single quantum tensor, we propose the workflow illustrated in Fig.~\ref{fig:optimization}.
The procedure starts with the optimization of the corresponding classical TTN (with classical tensors at the top) at a fixed bond dimension $\chi$.
The optimized TTN is then used to initialize the hTTN (transition at \textbf{($\ast$)}).
We first move the isometrization center of the hTTN to the quantum tensor.
All classical tensors in the hTTN are then initialized from the corresponding tensors in the TTN.
The quantum tensor is initialized by mapping the upper layer(s) of the TTN to a quantum circuit.
This results in a quantum circuit that consists of general two-qubit gates that are organized in a layered ladder topology~\cite{ran2020encoding,rudolph2023decomposition} (see Fig.~\ref{sfig:ladder_circuit}).
A ladder quantum circuit with $n$ layers approximately represents an MPS with bond dimension $\chi = 2^n$~\cite{rudolph2023decomposition}.
The quantum circuit is then extended with additional layers of two-qubit gates to increase the (virtual) bond dimension of the underlying hTTN to $\chi' > \chi$.
All gates are decomposed with the Cartan decomposition~\cite{helgason1979differential,vidal2004universal,vatan2004optimal}, resulting in 15 variational parameters per 2-qubit gate.
Each gate in the additional layers is initialized close to the identity by choosing small initial parameters (drawn from a Gaussian distribution with zero mean and standard deviation of 0.01) such that the state of the composed quantum circuit remains close to the pre-optimized one, while still evading known trainability issues when initializing gates as identities~\cite{funcke2021dimensional}.

Since the mapping of the classical tensors to a quantum circuit is approximate, the energy trajectory at \textbf{($\ast$)} is discontinuous.
However, this mapping already provides a significantly better initial point than a fully random parameter initialization~\cite{rudolph2023nature}.
Note that alternative quantum circuit topologies can also be used to represent the quantum tensor, e.g. a brick-wall circuit (see Fig.~\ref{sfig:brick_wall_circuit}).
As long as the circuit only consists of generic one- or two-qubit unitaries, the initialization strategy in Ref.~\cite{rudolph2023decomposition} can be safely employed.

Note that, after initializing the hTTN, the quantum tensor is the isometrization center and all classical matrices $\{P_l\}$ associated with it are equal to the identity matrix, \textit{i.e.}, $P_l = \mathbb{I}$.
Only after moving the isometrization center away from the quantum tensor the matrices will start to deviate from the identity.

The procedure continues with the sweep-based optimization of the hTN.
Each sweep starts with the optimization of the quantum tensor at the top (green region) and proceeds with the optimization of the classical tensors (blue region).
The optimization path through the hTTN is chosen to minimize the number of times the isometrization center is moved through the quantum tensor and, consequently, the quantum-computing overhead.
At the end of every sweep (at \textbf{($\bullet$)}) the isometrization center is moved back to the quantum tensor, to prepare it for the next iteration.
In general, the classical matrices $\{P_l\}$ associated with the quantum tensor will not result in identities anymore, and must be properly included in the quantum tensor optimization.
We propose the following options to handle them:

\begin{itemize}
    \item[(\textit{i})] \textit{No approximation:} 
    The quantum tensor is optimized by defining the loss function in Eq.~\eqref{eq:loss_with_penalty} without approximating the classical matrices $\{P_l\}$.
    Therefore, there is no discontinuity in the energy at \textbf{($\bullet$)}.
    However, satisfying the penalty term significantly increases the complexity of the optimization, and can even render it impossible.
    \item[(\textit{ii})] \textit{Projection to unitaries:} 
    As already observed in section~\ref{ssec:local_opt}, the loss function in Eq.~\eqref{eq:loss_with_penalty} reduces to Eq.~\eqref{eq:loss_without_penalty} if all $\{P_l\}$ are unitary.
    Therefore, to simplify the optimization, we project each $P_l$ to the closest unitary matrix (according to the Frobenius norm) before starting the optimization.
    The projection is realized by first calculating the SVD decomposition $P_l = U_l S_l V_l^\dagger$ and, then, replacing the Schmidt value matrix $S_l$ with the identity, $P_l' = U_l V_l^\dagger$.
    The energy jump at \textbf{($\bullet$)} depends on the magnitude of the non-unitary component of $\{P_l\}$.
    \item[(\textit{iii})] \textit{Re-initialization from best classical approximation:} 
    A third possibility is to neglect the penalty term in the loss function by re-initializing the quantum tensor from a classical TTN optimization at bond dimensions $\chi$. 
    Only the tensors included in the sub-network corresponding to the quantum tensor are optimized, while the rest of the TTN is fixed at the state reached after the preceding iterations. 
    The optimized tensors are then mapped to a quantum circuit and extended similarly to the initial mapping at point \textbf{($\ast$)} mentioned above.
    This again implies that all $P_l = \mathbb{I}$ and, therefore, the loss function without the penalty term (Eq.~\eqref{eq:loss_without_penalty}) can safely be used for optimizing the quantum tensor.
    The drawback of this option is that no information about the preceding optimization sweeps of the target quantum tensor is reused in the quantum tensor initialization.
    Therefore, the energy at \textbf{($\bullet$)} jumps above the dashed line.
    However, in all our tests we observed that this initialization is consistently more accurate than a fully random one.
\end{itemize}
In the following, we refer to these three strategies by their label (\textit{i}), (\textit{ii}), and (\textit{iii}).
We will compare their performances in Sec.~\ref{ssec:quantum_tensor_optimization}.

\subsection{Extension of the optimization workflow to \MakeLowercase{h}TN\MakeLowercase{s} with multiple quantum tensors}\label{sec:multiple_quantum_tensors}

Although we solely considered so far hTTNs with a single quantum tensor, the workflow introduced above is valid also for arbitrary loop-less hTNs, including those with multiple quantum tensors, as , e.g., the one depicted in Fig.~\ref{sfig:multiple_qtensors_ttn}. 
In this case, the handling of the $\{P_l\}$ matrices must not only be considered when starting a new sweep (i.e. at \textbf{($\bullet$)} in Fig.~\ref{fig:optimization}), but prior to the optimization of each quantum tensor.

\begin{figure}
    \centering
    \includegraphics[width=0.9\columnwidth]{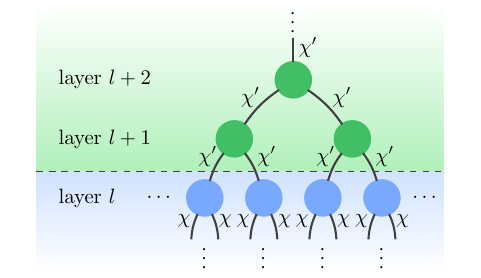}
    \caption{Section of a hTTN with interface between classical tensors and multiple quantum tensors.}
    \label{fig:multiple_qtensors_scaling}
\end{figure}

While the advantage of hTTNs with a single quantum tensor over their classical counterpart stems directly from the possibility of encoding states with a large virtual bond dimension, a different analysis is required when constructing hTTNs from multiple quantum tensors.
Consider, for instance, an hTTN with a classical-to-quantum interface between layers $l$ and $l+1$ at bond dimension $\chi'$.
The classical tensors in layer $l$ are of shape $(\chi, \chi, \chi')$, and all quantum tensors in layers $l' \geq l+1$ are of shape $(\chi', \chi', \chi')$, with $\chi < \chi' \leq \chi^2$.
The setup is sketched in figure~\ref{fig:multiple_qtensors_scaling}.
Note that, in the setup used in the previous examples, a single quantum tensor may represent multiple tensors of a classical TTN contracted together.
The \emph{virtual} bond dimension between these classical tensors depends on the number of layers used to encode the quantum tensor.
The setup shown in Figure~\ref{fig:multiple_qtensors_scaling} is intrinsically different because each quantum tensor in the hTTN after layer $l$ is associated with a single classical tensor in the corresponding TTN.
The bond dimension of the quantum tensors $\chi'$ cannot, therefore, be increased by adding layers of two-qubit gates to their quantum circuit representation.
Still, as we discuss in the following, hTTNs including multiple quantum tensors can in principle enhance purely classical TTNs.
In the first two rows of table~\ref{tab:scalings}, we list the computational complexity for storing and classically optimizing tensors of dimensions $(\chi, \chi, \chi')$ and $(\chi', \chi', \chi')$, respectively.
For simplicity, we use the worst-case scenario where $\chi'=\chi^2$.
In the following, we assume that $\chi$ has a value such that a $O(\chi^6)$ scaling is the highest we can classically handle. 
Then, the classical optimization of tensors in layer $l$ is feasible, whereas tensors in layers $l' \geq l+1$ cannot be optimized.
However, the optimization is possible when all tensors in layers $l' \geq l+1$ are quantum tensors, since the worst scaling of all quantum tensor related operations listed in table~\ref{tab:scalings} is $O(\chi^6)$.
This can be attributed to the fact that both, the link tomography required to move the isometrization center and the preprocessing of observables, can be done \textit{locally} on a single link of the quantum tensor.
Therefore, they merely scale with the link dimension $\chi^2$ instead of the tensor dimension $\chi^6$.
Here, we do not consider the complexity of the optimization of the quantum tensors.
If this computational step scales better than $O(\chi^8)$, then the overall protocol yields an advantage over its classical counterpart.

\begin{table}
    \centering
    \begin{tabular}{|c|c|c|c|}\hline
     Operation      & Scaling       & $(\chi, \chi, \chi^2)$ & $(\chi^2, \chi^2, \chi^2)$ \\\hline
     Storing        & $O(n)$        & $O(\chi^4)$   & $O(\chi^6)$     \\ 
     Optimization (classical) & $O(m^2)$      & $O(\chi^6)$~\cite{Diagonalization}   & $O(\chi^8)$     \\ 
     Link Tomography& $O(m^2)$ & -             &$ O(\chi^4)$     \\
     Observable preprocessing & $O(m^3)$        & -   & $O(\chi^6)$     \\ 
     \hline
    \end{tabular}
    \caption{
    Table showing the scaling for different operations applied to (h)TTNs at layer $l$, with tensors of shape $(\chi, \chi, \chi^2)$ and at layer $l'\geq l+1$ with tensors of shape $(\chi^2, \chi^2, \chi^2)$.
    With $n$ we denote the total number of elements in a tensor, and with $m$ the size of its leading
    dimension.
    }
    \label{tab:scalings}
\end{table}

\section{Results}
\label{sec:results}

We benchmark our proposed hTTN framework on the ground-state optimization of two different Hamiltonians: the transverse field Ising model and the Toric code.
The classical part of the hTTN calculations is based on the TTN implementation in Ref.~\cite{qtealeaves}.
All simulations of quantum computations are performed with an MPS simulator.
The bond dimension in the MPS simulator is not a major computational bottleneck for the system sizes considered below.
Therefore, all quantum operations are simulated in a noiseless setting and without any truncation of the bond dimension.
We optimize quantum tensors with the Variational Quantum Eigensolver (VQE)~\cite{peruzzo2014variational,kandala2017hardware} based on the L-BFGS-B optimizer~\cite{byrd1995limited} using a maximal number of 1000 iterations.

\subsection{The Ising model}
\label{ssec:ising_model}

The Hamiltonian of an $N$-site transverse field Ising model reads
\begin{equation}
  \hat{H} = -J \sum_{\langle i\,j \rangle} \sigma^x_i \sigma^x_j + h \sum_i \sigma^z_i \, ,
  \label{eq:ising_model}
\end{equation}
where $\langle i\,j \rangle$ denotes the sum over nearest neighbours.
In the following, we always consider the model at the critical point $|J| / |h| = 1$ and with periodic boundary conditions.
The minimal example for which a hTTN can be constructed and applied to the ground state search is a one-dimensional chain with $N=8$ sites, where the hTTN has two layers, as shown in Fig.~\ref{sfig:hybrid_ttn}.
We use this minimal system as a toy-model to compare the different strategies introduced in section~\ref{ssec:opt_workflow} for handling the classical matrices $\{P_l\}$ during the optimization of the quantum tensor.

\subsubsection{Benchmarking Quantum Tensor Optimization: the 8-site Ising Hamiltonian}
\label{ssec:quantum_tensor_optimization}

The hTTN setup depicted in Fig.~\ref{sfig:hybrid_ttn} yields an interface between classical and quantum tensors across links with bond dimension $\chi=4$.
In the following analysis, we will assume that this is the maximal bond dimension that can be achieved classically.
Therefore, we consider the ground state energy obtained with a $\chi = 4$ TTN as the classical reference energy.
We initialize the quantum tensor by mapping the top layer of the optimized $\chi=4$ TTN to two layers of a ladder circuit (see Fig.~\ref{sfig:ladder_circuit}), and then extend it with two additional layers. 
The quantum circuit obtained by varying the parameters of the two-qubit gates encodes a set of quantum states corresponding to MPSs with a bond dimension $\chi = 16$.
However, since this set does not fully cover the set of MPSs with $\chi = 16$, it is not guaranteed that the exact ground state (which is exactly represented by an MPS with $\chi = 16$ for $N=8$ sites) is contained in this set~\cite{rudolph2023decomposition}.
Nevertheless, the quantum circuit expressivity can be further increased by extending the circuit with more layers.

\begin{figure}
    \centering
    \subfloat[Hybrid optimization\label{sfig:optimization_strategies}]{
        \centering
        \includegraphics[width=0.9\columnwidth]{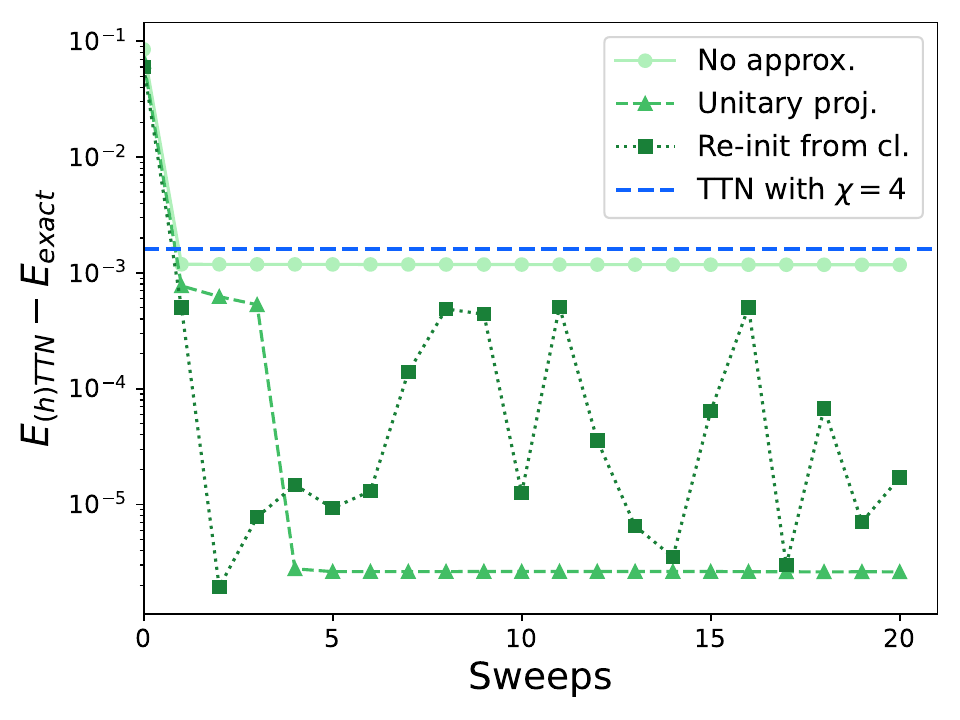}
    }\\
    \subfloat[Quantum tensor optimization\label{sfig:vqe_optimization}]{
        \centering
        \includegraphics[width=0.9\columnwidth]{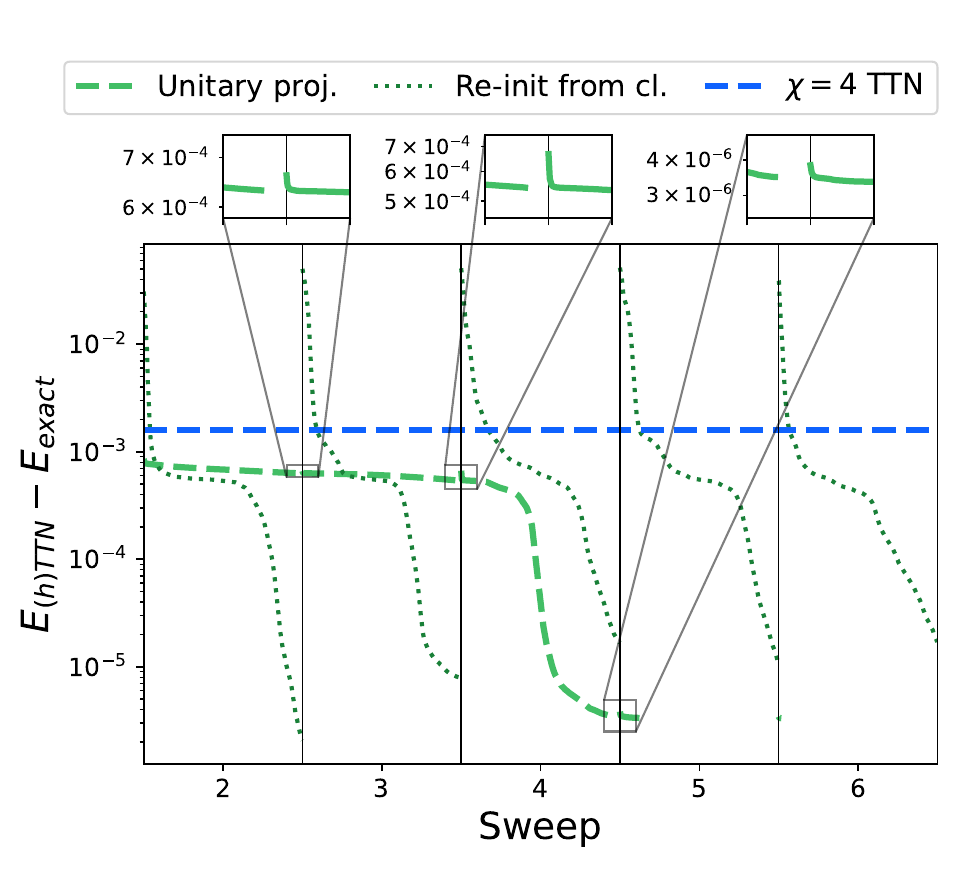}
    }
    \caption{
        Comparison between the different strategies introduced in section ~\ref{ssec:opt_workflow} for handling the classical matrices $\{P_l\}$ associated with the quantum tensor.
        (a) Ground-state energy calculated at the end of each sweep.
        For strategy (i) (No approx.), $\lambda = 1000$ is used as the scaling factor of the penalty term in Equation~\eqref{eq:loss_with_penalty}.
        (b) Ground-state energy evolution during the optimization of the quantum tensor in sweeps 2-6 when using either strategy (\textit{ii}) (green dashed line) or strategy (\textit{iii}) (green dotted line).
        The insets show a detailed view of the energy evolution for strategy (\textit{ii}) at the start of each optimization.
        Note that, for strategy (\textit{ii}) the optimization in sweeps 5 and 6 stops after a few iterations once it reaches convergence.
    }
\end{figure}

In the first sweep of the hybrid optimization, the classical matrices $\{P_l\}$ associated with the quantum tensor are equal to the identity (see Sec.~\ref{ssec:opt_workflow}).
Therefore, we can apply the VQE algorithm to the loss function in Eq.~\eqref{eq:loss_without_penalty} without including the penalty term to optimize the quantum tensor parameters.
Starting from the second sweep, the different strategies to handle the classical matrices $\{P_l\}$ result in different outcomes.
The results are shown in Fig.~\ref{sfig:optimization_strategies}.

Strategy~(\textit{i}) (circles on solid line) yields the worst convergence rate.
Accounting for the exact classical matrices relies on the loss function in Eq.~\eqref{eq:loss_with_penalty}, which includes the penalty term that enforces the quantum tensor normalization.
This penalty term appears to significantly increase the complexity of the optimization landscape, as the optimizer essentially fails to improve the energy beyond the result of the first sweep. 
The results are shown for $\lambda = 1000$, and equivalent results were also observed for other values of $\lambda$.
Both, strategy~(\textit{ii}) (triangles on dashed line) and (\textit{iii}) (squares on dotted line) yield an energy closer to the exact ground state energy.
The best achieved ground-state energy estimate is about 3 orders of magnitude more accurate than the energy obtained with the classical $\chi=4$ TTN.
We highlight that this simple experiment represents the first demonstration of a hTTN which improves the corresponding (artificially restricted) classical TTN.
The most noticeable difference between strategies~(\textit{ii}) and~(\textit{iii}) is that the former yields a much smoother convergence than the latter. 
The origin of this behaviour can be observed in Fig.~\ref{sfig:vqe_optimization}, which shows the evolution of the energy during the optimization of the quantum tensor in sweeps 2-6.
This figure highlights the key difference between strategy~(\textit{ii}) (green dashed lines) and strategy~(\textit{iii}) (green dotted lines), namely the energy at the start of the optimization.
The energy obtained from the optimized classical tensors in strategy (\textit{iii}) is at most as good as the energy of the $\chi = 4$ TTN (blue dashed line).
Since the mapping from the classical tensors to the quantum circuit is inherently approximate, the initial energy for strategy (\textit{iii}) will always be higher than the energy obtained with the $\chi = 4$ TTN, and lies therefore above the blue dashed line.
Conversely, for strategy~(\textit{ii}) the initial energy depends on how close the classical matrices $\{P_l\}$ are to being unitary.
For this test case, these matrices are nearly unitary and, therefore, there is barely no energy jump between consecutive sweeps (see insets in Fig.~\ref{sfig:vqe_optimization}).
This also directly influences the number of optimization iterations in each sweep.
With strategy (\textit{iii}), many iterations are required to converge back to the energy obtained in the previous sweep before any improvement can be made.
In some cases, the pre-defined maximal number of iterations is even not sufficient to reach an energy better than the previous one.
This leads to the bumps observed for the curve obtained with strategy (\textit{iii}) reported in Fig.~\ref{sfig:optimization_strategies}.
Conversely, for strategy~(\textit{ii}) only small jumps in the energy convergence are observed.
In fact, less iterations are required in this case to converge back to the energy of the previous sweep, and most of the iterations are used to improve upon it.

Based on these observations, we therefore will use strategy~(\textit{ii}) -- namely projecting the classical matrices $\{P_l\}$ to the closest unitary -- in all subsequent simulations.

\subsubsection{Targeting larger systems: the 16-site Ising Hamiltonian}

\begin{figure}
    \centering
    \includegraphics[width=.9\columnwidth]{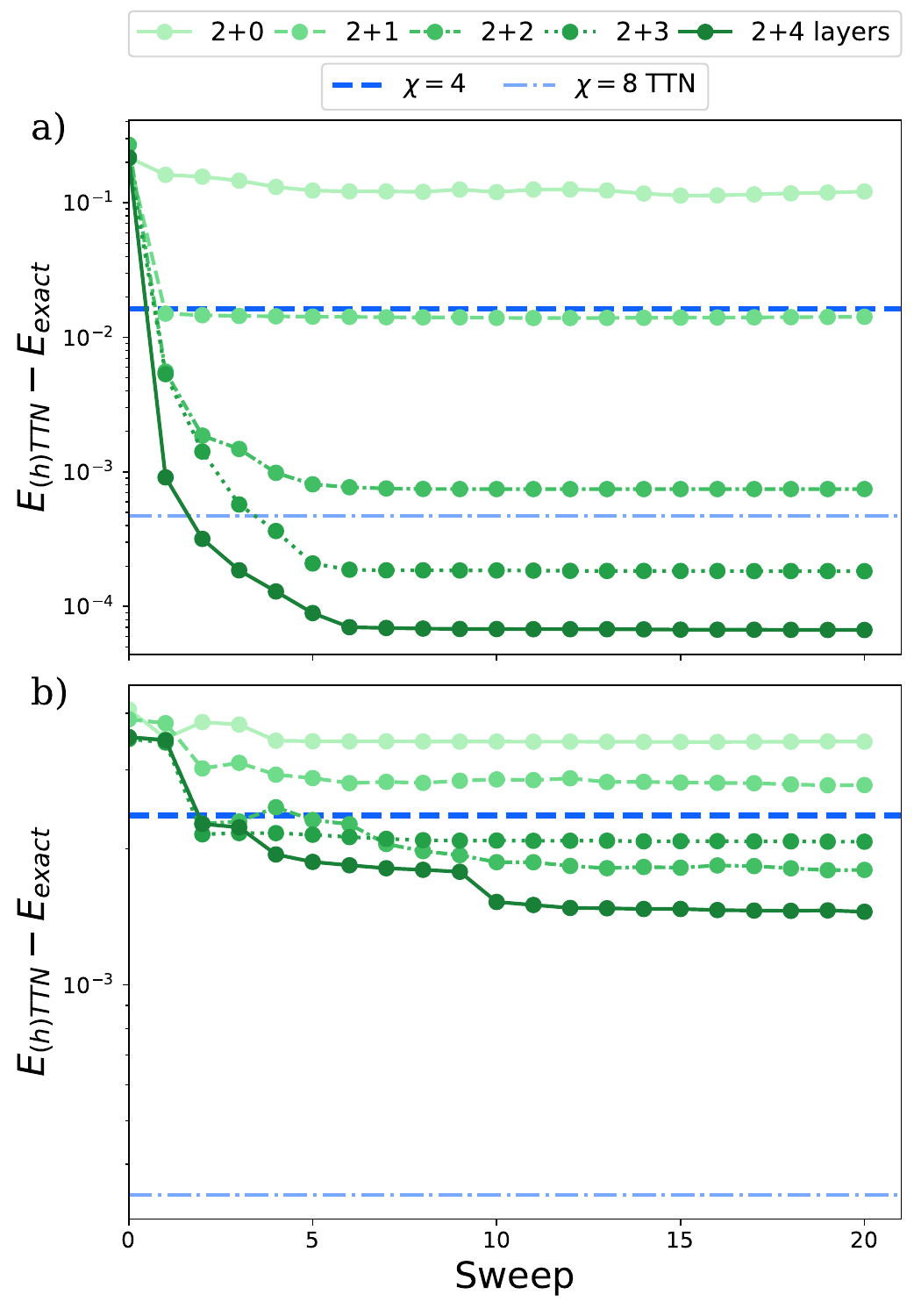}
    \caption{
    Optimization of one- (a) and two-dimensional (b) 16-site Ising model with a hTTN. 
    The initial point for all hTTN optimizations is an optimized classical TTN with bond dimension $\chi = 4$.
    In all experiments, the quantum tensor is represented with a quantum circuit consisting of $m + e$ layers of the ladder quantum circuit, where $m$ is the number of layers used for the mapping of the optimized classical TTN, and $e$ is the number of layers used to extend the quantum circuit in order to increase its (virtual) bond dimension.
    Reference energies obtained with a classical TTN of bond dimension $\chi = 4, 8$ are indicated with horizontal blue lines.}
    \label{fig:IM_16}
\end{figure}

We will now scale up the size of our benchmarks and apply the proposed hTN approach to optimize the ground state of the 16-site Ising Hamiltonian, both in one dimension and on a $4\times4$ square lattice.
We apply in both cases periodic boundary conditions.
The classical binary TTN required to represent a wave function of the 16-site Ising model consists of three layers of classical tensors.
Based on this topology, we construct a two-layer hTTN with a single quantum tensor that represents the two topmost layers of the classical TTN.
The interface between the classical tensors and the quantum tensor is of bond dimension $\chi = 4$.
Therefore, the quantum tensor corresponds to a quantum state defined on 16 qubits (2 qubits per leg).
As above, we construct the quantum circuits encoding the quantum tensor by mapping the pre-optimized classical tensors to an appropriate number of layers of the ladder quantum circuit (see Fig.~\ref{sfig:ladder_circuit}).
The results of the optimizations are shown in Fig.~\ref{fig:IM_16}a for the one-dimensional and Fig.~\ref{fig:IM_16}b for the two-dimensional Ising model, respectively.

Both figures show the ground-state energy estimate obtained at the end of each sweep for different numbers of layers in the quantum circuit representation of the quantum tensor.
A quantum circuit with $m+e$ layers consists of $m$ layers resulting from for the initial mapping of the reference TTN state (dashed blue line) to a quantum circuit, and $e$ layers for extending it to a higher (virtual) bond dimensions (see Fig.~\ref{fig:quantum_circuits}).
In all cases, we use $m=2$ and consider values for $e$ between 0 and 4.
The blue horizontal lines indicate reference energies obtained from a classical TTN optimization with bond dimensions $\chi = 4$, and $8$.
The starting point for the hTTN optimizations, to which the quantum-enhanced version should be compared, is the line corresponding to the $\chi = 4$ TTN (blue dashed line).
An improvement over this classical reference is already obtained with a $2+1$ layer quantum circuit for the one-dimensional system, and with a $2+2$ layer quantum circuit for the two-dimensional system.
As expected, increasing the number of layers in the quantum circuit results in consistently better energy estimates.

It is noteworthy that, for a given number of layers, the energy improvement over the classical reference is significantly larger for the one-dimensional Hamiltonian than for the two-dimensional one.
Indeed, whereas in the one-dimensional case it is possible to improve over the classical energy by more than two orders of magnitude with the $2 + 4$ layer quantum circuit, the improvement is reduced to only a factor of 2 in the two-dimensional case.
The only difference between these two systems lies in the qubit representation of the Hamiltonian and, therefore, in the specific observable measured to estimate the energy.
This results in an more complex entanglement structure in the ground state wave function in the two-dimensional model than in the one-dimensional one.
It is, therefore, not surprising that the ladder quantum circuits, which are directly inspired by one-dimensional MPS structures, are better suited to one-dimensional Hamiltonians.
In the next Section we investigate options to increase the expressivity of the quantum circuit ansatz beyond MPS-based topologies.

\subsubsection{Improved Circuit Topology Design for Quantum Tensors}

\begin{figure}
    \centering
    \includegraphics[width=0.9\columnwidth]{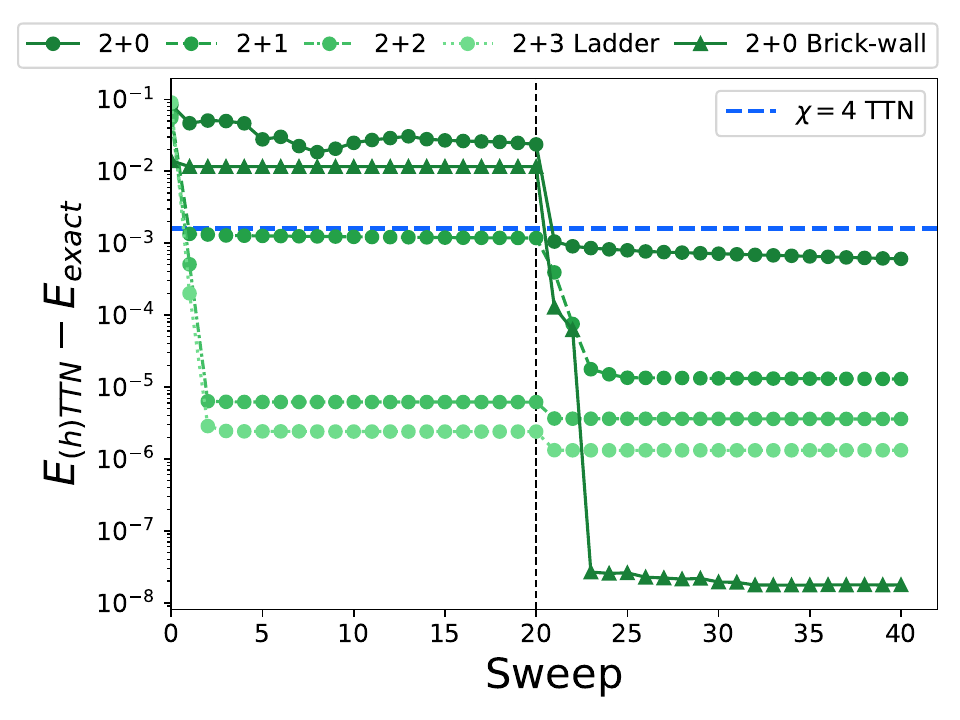}
    \caption{
        Ground-state energy trajectory obtained for the one-dimensional 8-site Ising model obtained by changing the circuit topology during the optimization for the ladder circuit (circles) and the brick-wall circuit (triangles).
        The initial point for all hTTN optimizations is an optimized classical TTN with bond dimension $\chi = 4$.
        The horizontal blue line shows the reference energy of the classical $\chi = 4$ TTN.
        After the 20$^{\text{th}}$ sweep (black dashed vertical line), an additional gate is added to the quantum circuit connecting the first and last qubit for each layer of two-qubit gates.
    }
    \label{fig:IM_08_circuit_topology}
\end{figure}

We now move back to the one-dimensional 8-site Ising model introduced above (Sec.~\ref{ssec:quantum_tensor_optimization}).
First, we consider a small variation of the ladder circuit topology of Fig.~\ref{sfig:ladder_circuit}.
Specifically, we add to each layer of the quantum circuit an additional two-qubit gate connecting the first and last qubit.
This effectively promotes the circuit to a periodic one, which matches more closely the topology of the model system under study.
We isolate the effect of these additional gates by means of the following experiment: we first optimize the ground state using a hTTN where the quantum tensor is represented by the conventional ladder quantum circuit.
We then insert the additional gates (initialized as identities) after 20 sweeps, and finally proceed with the optimization for another 20 sweeps.
The results are shown in Fig.~\ref{fig:IM_08_circuit_topology}.
The additional gates have essentially the same effect as adding a full additional layer to the circuit.
For shallow circuits ($2+0$ and $2+1$ layers), adding the gates substantially improves the energy.
Conversely, for deeper circuits ($2+2$ and $2+3$ layers) the circuit is already expressive enough, such that adding more gates does not improve significantly the ground-state energy estimate.
On current quantum processors, shallower quantum circuits are easier to realize.
Therefore, in the perspective of implementing our approach on quantum computer, it is encouraging to observe that a few well-placed and physically motivated  gates are sufficient to largely improve the accuracy of our approach.

To generalise even further, we consider a circuit design, which is not derived from the MPS-to-quantum-circuit mapping applied above, namely a quantum circuit with a $2+0$ brick-wall structure (see Fig.~\ref{sfig:brick_wall_circuit}).
The number of gates is the same as for the $2+0$ ladder circuit used above (14 without and 16 with additional gates).
To assess the quality of this design, we repeat the same numerical experiment described above: we add generic two-qubit gates connecting the first and last qubit after 20 sweeps of optimization and then continue the optimization for another 20 sweeps.
The initial parameters for the gates are obtained by applying the fidelity optimization procedure introduced in Ref.~\cite{rudolph2023decomposition}.
The results are reported in Fig.~\ref{fig:IM_08_circuit_topology}, where we can observe a qualitative behaviour consistent with the previous analysis of ladder circuits.
We note that the brick-wall circuit is better suited to represent the target ground state compared to ladder type circuits.
However, for larger system sizes, a generic brick-wall circuit may suffer from trainability issues~\cite{thanasilp2023subtleties}.

In summary, we believe that the construction of the most effective quantum circuits will employ a certain level of model-inspired design, as, e.g., the use of additional ``wrapping'' gates to improve the description of periodic systems.
The optimal combination of classical and quantum tensors will enable expressivity and trainability superior to the ones achievable by either purely classical or quantum TNs.

\subsubsection{Effect of statistical noise on the hTTN optimization}
\label{sec:noisy_experiments}
\begin{figure}
    \centering
    \includegraphics[width=0.9\columnwidth]{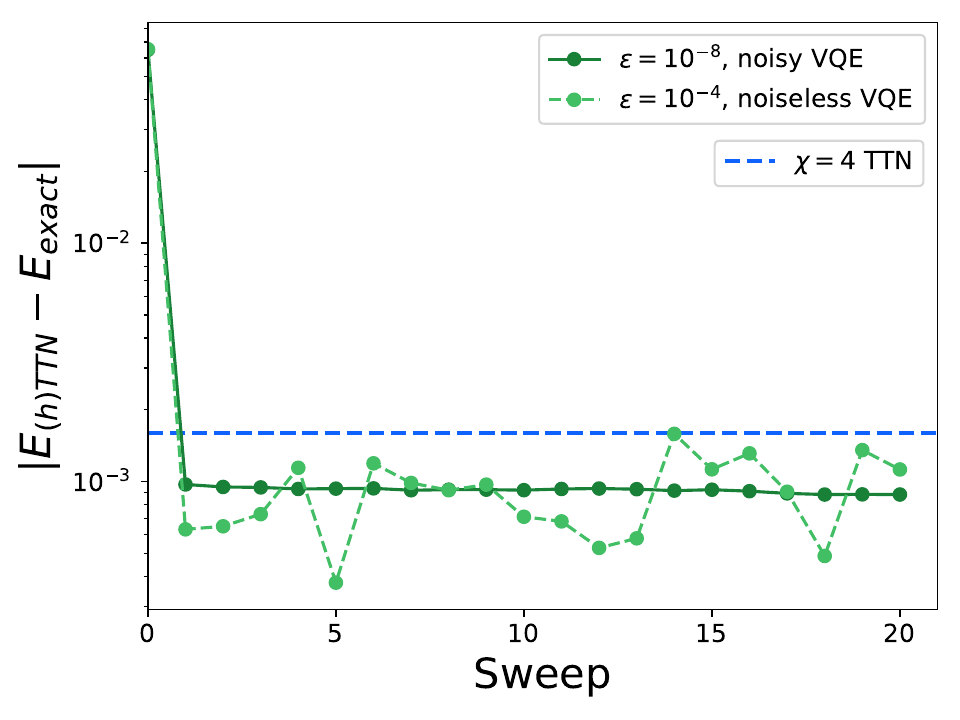}
    \caption{Optimization of hTTN for 8-site Ising model in the presence of statistical noise in all operations involving the quantum tensor (solid, dark green line), and all operations except the VQE optimization of the quantum tensor (dashed, light green line).
    The initial point of the hTTN optimizations is an optimized classical TTN with bond dimension $\chi = 4$.
    The horizontal blue line shows the reference energy of the classical $\chi = 4$ TTN.}
    \label{fig:statistical_noise}
\end{figure}

In the following, we investigate the effect of statistical noise when estimating expectation values with a quantum tensor.
When considering the conventional way of evaluating expectation values (i.e. contracting all legs of a quantum tensor) as $\langle \hat{O} \rangle = \bra{\psi} \hat{O} \ket{\psi}$, we can approximately account for statistical noise by adding a normally distributed random variable to the noiseless expectation value according to the distribution $\eta_\epsilon \sim \mathcal{N}(0, \epsilon)$, where $\epsilon$ measures the strength of the statistical fluctuations.
For the open link contraction, which is evaluated during the implicit isometrization (see section~\ref{ssec:implicit_isometrization}) or the calculation of the effective Hamiltonian (see section~\ref{ssec:local_eff_ham}), emulating the statistical noise becomes less straightforward. 
Due to the architecture of the applied hTTN, we limit ourselves to the case where the open link is a quantum index (see equation~\eqref{eq:measurement_matrix_quantum}).
Similar considerations also hold for an open classical index (see equation~\eqref{eq:measurement_matrix_classical}).
On a quantum device, we evaluate the open-link contraction according to equation~\eqref{eq:open_link_reconstruction}. 
For each expectation value $E(\sigma) = \bra{\psi} \sigma \otimes \hat{O} \ket{\psi}$ appearing in the sum, the statistical noise is approximated as described above.
Each $E(\sigma)$ factor is replaced with its noisy counterpart $E(\sigma)_\epsilon = E(\sigma) + \eta^\sigma_\epsilon$, where $\eta^\sigma_\epsilon \sim \mathcal{N}(0, \epsilon)$ is the statistical noise added to the expectation value of the Pauli string $\sigma$.
This changes equation~\eqref{eq:open_link_reconstruction} to
\begin{align}\label{eq:noisy_exp_val}
    \tilde{M} &= \frac{1}{\chi} \sum_{\sigma} (-1)^{y(\sigma)} (E(\sigma) + \eta^{\sigma}_\epsilon) \sigma \\
              &= M + \frac{1}{\chi} \sum_\sigma (-1)^{y(\sigma)} \eta^\sigma_\epsilon \, \sigma = M + M_\epsilon \,,
\end{align}
where $M$ is the noise-less evaluation of the open-link contraction, and $M_\epsilon$ is the additional noisy operator added whenever an open-link contraction is evaluated for a quantum tensor. \\
We consider two experimental setups for noisy simulations of the hTTN optimization of the 8-site Ising model. 
The first one considers statistical noise for all operations including the quantum tensor.
The second one performs the quantum tensor optimization (i.e. the VQE) without statistical noise, and only considers it for the implicit isometrization and the calculation of the effective Hamiltonian.
We use a $\chi = 4$ TTN as the initial point for the hTTN, and a $2+2$-layer ladder quantum circuit to represent the quantum tensor.
In both setups, we identify the highest amount of the statistical noise $\epsilon$ that still results in energies below the energy obtained with a $\chi = 4$ classical TTN.
The results are shown in figure~\ref{fig:statistical_noise}.
Note that the energy scale now is the absolute value of the deviation from the exact ground state energy, since the statistical noise may violate the variational principle.
As for all the experiments reported above, the optimizer used in the VQE routine is BFGS, which performed better than all other tested alternatives (even SPSA~\cite{spall1992spsa}, the recommended optimizer in a noisy setting).
In the first case, where the VQE includes the statistical noise, the noise threshold lies at $\epsilon = 10^{-8}$, which would likely result in a large measurement overhead on a real quantum processor. However, when statistical noise in the VQE component is neglected, the noise threshold increases to a more manageable value of $\epsilon = 10^{-4}$.
This clearly indicates that, in the presence of statistical noise, a purely variational optimization of the quantum tensors will likely represent a severe bottleneck. 
This strengthens the need for improved quantum optimization strategies (e.g., local imaginary-time or adiabatic evolution) and for exploring hTTN architectures composed of fully-parametric classical tensors and few-parameters (or even, ideally, non-parametric) quantum circuits.

We expect that a similar error threshold on the statistical fluctuations would also apply to larger systems to ensure that the classical part of the computation converges properly.
However, the cost to reach that error threshold in the estimation of the relevant quantum mechanical observables will typically grow.
More specifically, the prefactor to the Monte Carlo-like $1/\sqrt{S}$ scaling (where $S$ is the number of measurement shots) typical of neat-term quantum computing approaches is, in general, at least polynomial in the system size.

Optimized measurement techniques like advanced Pauli groupings, classical shadows or, more generally, informationally complete (IC) quantum measurements could be used to address this task in an effective way. It is worth noticing here that such protocols have recently been realized at the scale of 90+ qubits on superconducting quantum processors~\cite{fischer2024dynamical}.
These methods would allow us to efficiently estimate a large number of observable with a reasonable sampling cost, provided that, e.g., some degree of locality is preserved in the measured operators.
Interestingly, our proposed approach can readily be improved, with respect to this aspect. 
For instance, one could focus on designing the classical TN part to limit the growth of the number and of the support of the transformed operators to be measured on the quantum circuit, e.g., by controlling the non-stabilizerness and bond dimension of the induced transformation, particularly if the original Hamiltonian is already $k$-local.

\subsection{Application to the Toric Code Hamiltonian}
\begin{figure}
    \centering
    \includegraphics[width=0.7\columnwidth]{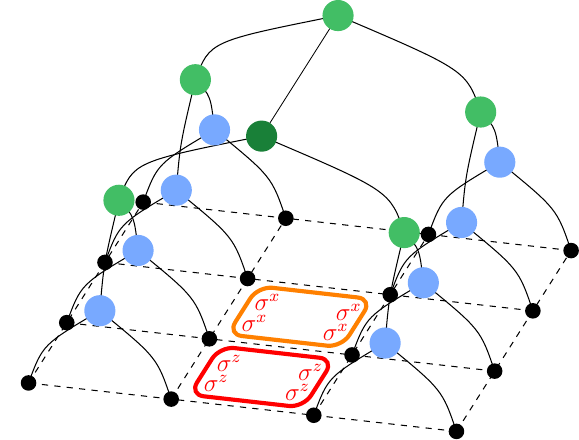}
    \caption{Toric code on a $4\times4$ lattice with the hTTN represented on top. The classical tensors of dimension
    $(2, 2, 4)$ are represented in blue, while the quantum tensors of dimension $(4, 4, 4)$ in green. The dark green
    tensor is the isometry center.}
    \label{fig:toric}
\end{figure}

In the following, we perform ground state optimizations for the Toric code Hamiltonian on a $4\times4$ lattice using a hTTN with multiple quantum tensors in the two highest layers of the tree, as shown in Fig.~\ref{fig:toric}.

The periodic $N$-site two-dimensional Toric code~\cite{kitaev2003} Hamiltonian is described by
\begin{equation}
    \hat{H}_T = \sum_{p_e} A_e + \sum_{p_o} B_o , \quad
    A_e=\begin{matrix}
      \sigma^{x}_{\ulcorner} & \sigma^{x}_{\urcorner}\\
      \sigma^{x}_{\llcorner} & \sigma^{x}_{\lrcorner}
    \end{matrix},\quad
    B_o=\begin{matrix}
      \sigma^{z}_{\ulcorner} & \sigma^{z}_{\urcorner}\\
      \sigma^{z}_{\llcorner} & \sigma^{z}_{\lrcorner}
    \end{matrix},
\end{equation}
where the index $p_e$ runs over the even plaquettes of the system and $p_o$ over the odd plaquettes.
The ground state of the Toric code is maximally entangled and therefore it is a very complex target for tensor networks simulations \cite{hermanns2017entanglement}.
Since the exact ground state energy is known exactly, this system represents an ideal benchmark for our proposed hTTN optimization algorithm.

Following the reasoning of Sec.~\ref{sec:multiple_quantum_tensors}, we take as the largest classical bond dimension the value of $\chi=2$ with an interface dimension of $\chi'=4$.
These conditions will allow us to validate our approach while keeping the classical and quantum resources within reasonable bounds.  
Consequently, the resulting hTTN (illustrated in Fig.~\ref{fig:toric}) is composed of a bottom layer with classical tensors of dimensions $(2, 2, 4)$ and two upper layers with quantum tensors of dimensions $(4, 4, 4)$.
All quantum tensor are represented by a $6$-qubit brick-wall quantum circuit with $n_{layers}=3$ layers. 
Notice that in this case, differently from the Ising model, we cannot straightforwardly initialize the quantum tensors from their classical counterparts, since we do not have access to an equivalent optimized classical tensor of shape $(4, 4, 4)$. 
We thus initialize the quantum tensors with random parameters.
As for the Ising chain, prior to the optimization of the quantum tensor we again use strategy (\textit{ii}) -- the projection to the closest unitary -- for the handling of the classical matrices associated with a given quantum tensor. 
To limit the computational cost, we restrict to 100 iterations for the VQE optimization of the quantum tensors.

\begin{figure}
    \centering
    \subfloat[Two dimensional ($4\times4$ lattice)\label{sfig:toric4x4}]{
        \centering
        \includegraphics[width=0.9\columnwidth]{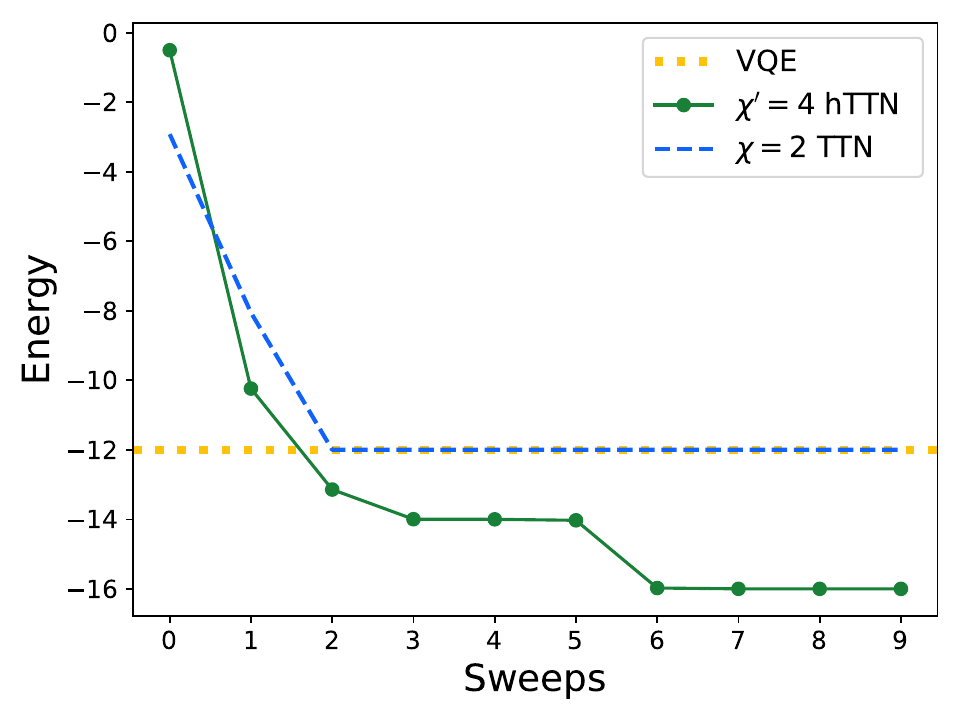}
    }\\
    \subfloat[Two dimensional ($8\times8$ lattice)\label{sfig:toric8x8}]{
        \centering
        \includegraphics[width=0.9\columnwidth]{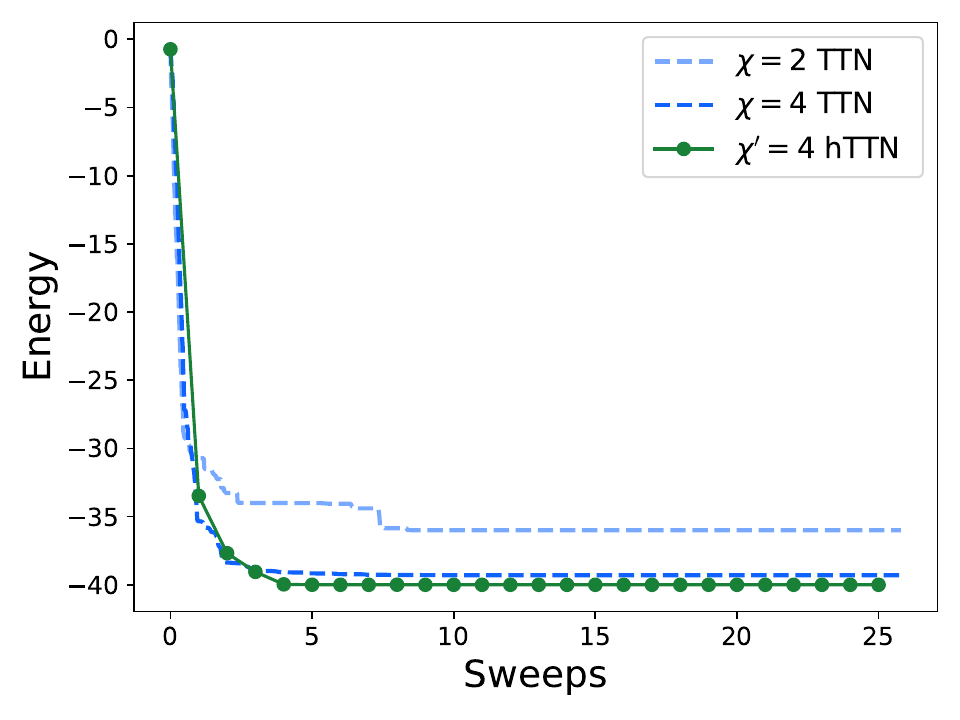}
    }
    \caption{
        (a) Ground state search for the $4\times4$ Toric code using a hTTN that contains multiple quantum tensors (dots on green line), and reference experiments with a classical $\chi = 2$ TTN (blue, dashed line), and a 16-qubit VQE optimization of the full system (orange, dotted line) using a ladder quantum circuit with $n_{layers}=6$ layers. 
        The figure shows the best optimization run from 100 realizations of the experiment for each of the methods.
        For the 16-qubit VQE optimization, we only report the final energy.
        (b) Ground state search for the $8\times8$ Toric code using a hTTN that contains multiple quantum tensors (dots on green line), and reference experiments with a classical $\chi = 2$ TTN (light blue, dashed line) and a $\chi=4$ TTN (dark blue, dashed line).
    }
    \label{fig:toric_results}
\end{figure}

The energy optimization with this hTTN setup is shown in Fig.~\ref{sfig:toric4x4} (dots on green, solid line) and compared to a classical $\chi=2$ TTN (blue, dashed line) and to a 16-qubit VQE optimization of the full system (orange, dotted line). In the full-system VQE optimization, the applied variational quantum circuit is the ladder circuit with $n_{layers}=6$.
We also report as a reference the final energy of the VQE simulation, which corresponds to the lowest value reached by the simulation, as the optimization reached convergence.
Each result is the best realization over $100$ runs of the respective experiment.
Notice that only the hTTN approach is able to retrieve the exact ground state energy of $E_{\text{exact}} = -16$.

After successfully recovering the ground state energy for the $4\times 4$ lattice, we scale up to the $8\times 8$ Toric code system.  
It is worth mentioning that this system -- despite its modest size -- can hardly be solved by exact classical methods, while a direct optimization with a bare VQE on 64 qubits could become extremely challenging.
On the other hand, by means of the proposed hTTN algorithm we can aim to solve the same problem using $6$-qubit brick-wall quantum circuits with $n_{layers}=3$ with an interface of bond dimension $\chi'=4$.
Remarkably, as shown in Fig.~\ref{sfig:toric8x8} the hTTN with $\chi'=4$ (dots on green, solid line) not only outperforms the $\chi=2$ TTN (light blue dashed line) but recovers an energy lower than the $\chi=4$ TTN (dark blue dashed line).
While we leave the investigation of higher bond dimensions $\chi'$ to future works, these results, together with the ones reported above, clearly suggest that hybrid approaches can be successfully employed to extend the reach of state-of-the-art computational techniques leveraging quantum resources.

\section{Conclusions}
\label{sec:conclusion}

Combining classical tensor networks and quantum computing is a promising route for studying quantum many-body systems.
Hybrid tensor networks~\cite{yuan2021quantum} realize this idea by encoding tensors in quantum states prepared on quantum computers.
In this work, we investigated the potential of this approach for the encoding of large dimensional tensors in a TN in order to reach bond dimensions which are classically unfeasible.
To this end, we introduced an scalable optimization scheme for loop-less hTNs inspired by algorithms originally designed for classical TTNs.
The global optimization of an hTN is split into a sequence of local optimizations of the individual tensors, which are optimized variationally.
Our new algorithm is largely more scalable than the global optimization techniques that have been proposed so far~\cite{yuan2021quantum}, which become unfeasible for large hTNs.
In this study, we have successfully applied the proposed local optimization algorithm to the ground state search of the 8-, 16- and $4\times4$-site transverse field Ising model and the $4\times4$- and $8\times8$-site Toric code.
In all investigated instances, our simulations showed that hTNs can enhance, sometimes even significantly, the accuracy of the ground-state energy estimate compared to a corresponding classical TNs calculation whose bond dimension is (artificially) upper bounded.
Our results indicate that such behaviour may potentially extend to problem instances that can be hardly targeted by classical TNs.

Despite these promising preliminary results, the proposed framework still requires further improvements before achieving regimes of practical relevance~\cite{Kim2023_QuantumUtility}.
First, the VQE-like optimization scheme applied to quantum tensors may not scale to more than a few tens of qubits, due -- for instance -- to known issues about the trainability of large parameterized quantum circuits~\cite{mcclean2018barren,wang2021noise,thanasilp2023subtleties}.
Other optimization strategies, e.g. based on sweep-like optimization~\cite{slattery2022unitary} or by optimizing the individual quantum tensors with quantum imaginary time-evolution~\cite{lin2021real}, may be required to scale the optimization up to a larger number of qubits.
Another important bottleneck is represented by the number of measurements required to estimate expectation values on quantum tensors, e.g., when measuring the energy or during the construction of local effective Hamiltonians.
Indeed, the number of Pauli strings obtained from the decomposition of the effective Hamiltonian $\hat{H}_{\text{eff}}$ towards the top of a hTTN can easily become large even for relatively sparse lattice spin Hamiltonians, due to repeated contractions with classical tensors.
As we discuss in Appendix~\ref{app:local_diag_of_obs}, one can exploit the tensor product structure of the considered observables to trade the number of measurements for an increased quantum circuit depth.
In addition, the number of measurements may be further reduced by applying approximate tomography protocols~\cite{kaznady2009numerical,gross2010quantum,aaronson2018shadow,Torlai2023_TNS-Tomography,Akhtar2023_MPS-Tomography}, IC-POVMs~\cite{acharya2021shadow,garcia2021learning,fischer2022ancilla,fischer2024dual}, or possibly through sparse approximations of the Pauli representation of the effective Hamiltonian.

While our work mainly focused on a noise-free scenario, in Section~\ref{sec:noisy_experiments} we addressed the problem of statistical fluctuations.
We have shown that the open-link contractions involving quantum tensors -- the key step of our protocol -- is resilient to shot noise.
Importantly, our proposed hybrid optimization framework can be straightforwardly combined with optimized measurement techniques to achieve a similar resilience also for larger systems.
The optimization of the quantum tensor is, instead, less robust.
This is, however, not surprising, as the VQE optimization is known to suffer from similar limitations.
As previously mentioned, these results further strengthen the necessity for alternative and improved quantum optimization strategies.

Even though so far we focused on defining a general hTN framework, its efficiency can largely be improved by tailoring the network architecture and the quantum circuit Ans\"atze towards specific application targets.
Moreover, our hTTN toolkit can be readily extended to dynamical simulations, for instance by combining classical methods based on the time-dependent variational principle for TTN states~\cite{Haegeman2011_TDVP,Haegeman2016_TDVP,Bauernfeind_2020} with the corresponding quantum variational real-time evolution method~\cite{Benjamin2017_VarRQTE,Miessen2022_QuantumDynamics-Review}.

Finally, it is worth mentioning that our framework would also be applicable in combination with future fault tolerant quantum architectures by leveraging quantum phase estimation-inspired subroutines, to obtain precise estimates of complex observables at very modest sampling costs~\cite{knill2007optimal}. 
Interestingly, intermediate solutions, potentially viable in an early fault-tolerant setting, are also known~\cite{wang2021minimizing}.
In this scenario, hybrid TNs could be used as a strategy to embed quantum wave functions of strongly-correlated cores within a more loosely correlated classical environment.

\emph{Acknowledgments $-$} 
We thank Vincenzo Savona for helpful discussions.
This research was supported by the NCCR MARVEL,
a National Centre of Competence in Research, funded
by the Swiss National Science Foundation (grant number 205602). IBM, the IBM logo, and ibm.com are trademarks
of International Business Machines Corp., registered in
many jurisdictions worldwide. Other product and service names might be trademarks of IBM or other companies. The current list of IBM trademarks is available at
\url{https://www.ibm.com/legal/copytrade}.
We acknowledge financial support
from the Italian Ministry of University and Research (MUR) via PRIN2022 project TANQU, and the Departments of Excellence grant 2023-2027 Quantum Frontiers;
from the European Union via
H2020 projects EuRyQa, via the NextGenerationEU project CN00000013 - Italian Research Center on HPC, Big Data and Quantum Computing (ICSC), and from the World Class Research Infrastructure $-$ Quantum Computing
and Simulation Center (QCSC) of Padova University.
G.M. is partially supported by the Italian funding within the ``Budget MUR - Dipartimenti di Eccellenza 2023-2027" (Law 232, 11 December 2016) - Quantum Sensing and Modelling for One-Health (QuaSiModO), by UNIBA through the 2023-UNBACLE-0244025 grant. 
The authors thank the QC4HEP Working Group for discussions.

\bibliographystyle{apsrev4-2}
\bibliography{references}

\appendix

\section{Open quantum link contraction}
\label{app:open_link_contraction}

Ref.~\cite{yuan2021quantum} describes how to perform the measurement of an $k$ site observable with an $(k+1)$-rank tensor that is represented by a $(k+1)$-partite quantum state.
The expression to be evaluated is given in Eq.~\eqref{eq:measurement_matrix_quantum}.
Assuming that the open index is represented by a single qubit, this expectation value can be evaluated by measuring the expectation value of the operators $\sigma = \{\mathbb{I}, X, Y, Z\}$
\begin{equation}
    E(\sigma) = \langle \psi | \sigma \otimes O_1 \otimes O_2 \otimes \cdots \otimes O_k | \psi \rangle \,,
\end{equation}
where $\mathbb{I}$ is the $2\times2$ identity matrix and $X, Y, Z$ are the Pauli matrices.
The measurement matrix $M$ can then be reconstructed from the $E(\sigma)$ as
\begin{equation}
    M = \frac{1}{2} \left(E(\mathbb{I}) \mathbb{I} + E(X) X - E(Y) Y + E(Z) Z \right) \,.
\end{equation}
This process can be generalized to a quantum link that is implemented with an arbitrary number of qubits $n_{\text{qubits}} > 1$. 
The corresponding dimension of the link is then $\chi = 2^{n_{\text{qubits}}}$.
The procedure described above can be straightforwardly generalized by measuring all operators in the Pauli basis $\sigma \in \{\mathbb{I}, X, Y, Z\}^{\otimes n_{\text{qubits}}}$.
The measurement matrix $M$ can be reconstructed from the expectation values $E(\sigma)$ as follows
\begin{equation}\label{eq:open_link_reconstruction}
    M = \frac{1}{\chi} \sum_{\sigma} (-1)^{y(\sigma)} E(\sigma) \sigma \,,
\end{equation}
where $y(\sigma)$ is the number of $Y$ operators in the Pauli string $\sigma$.

\section{Positive semi-definiteness of measurement matrix}
\label{app:positive_semi_definiteness}
\begin{figure}
    \centering
    \includegraphics[width=0.35\columnwidth]{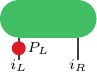}
    \caption{
        Quantum tensor with two legs.
        The leg on the left is connected to an arbitrary classical matrix $P_L$.
    }
    \label{fig:semidef}
\end{figure}

In this section, we demonstrate that the measurement matrix $M$ in Eq.~\eqref{eq:measurement_matrix_classical} and~\eqref{eq:measurement_matrix_quantum} is positive semi-definite if $O = \mathbb{I}$.
The matrix $M$ is, in principle, just a reduced density matrix of the quantum state(s) of the quantum tensor, and therefore positive semi-definite by construction. However, since we extend the quantum tensor with a set of classical matrices associated with its legs, or the legs can either be of classical or quantum nature, we proof here, that the matrix $M$ is also positive semi-definite in this case.
First, we consider the special case where we assume that the quantum tensor has two quantum indices, and a classical matrix $P_l$ contracted with the left index (see illustration in Fig.~\ref{fig:semidef}). 
The quantum tensor is the current isometrization center and we evaluate the open-link contraction with respect to the right index $i_R$.
We represent the quantum tensor as
\begin{equation}
    \ket{\psi} = \sum_{i_L i_R} \psi_{i_L i_R} \ket{i_L} \otimes \ket{i_R} \,,
\end{equation}
and the matrix on the left index as
\begin{equation}
    P_L = \sum_{ij} (P_L)_{ij} \ketbra{i}{j} \,.
\end{equation}
Plugging these expressions into Eq.~\eqref{eq:measurement_matrix_quantum} we obtain
\begin{align}
    M^{n' n} & = \bra{\psi} P_L^{\dagger} P_L \otimes (\ketbra{n'}{n}) \ket{\psi} \nonumber \\
             & = \sum_{i_L' i_L} \psi^*_{i_L' n'} (P_L^{\dagger} P_L)_{i_L' i_L} \psi_{i_L n} \nonumber \\
             & = \sum_{i_L' i_L} \sum_{i} \psi^*_{i_L' n'} (P_L^{\dagger})_{i_L' i} (P_L)_{i i_L} \psi_{i_L n} \nonumber \\
             & = \sum_{i} \left(\sum_{i_L'} \psi^*_{i_L' n'} (P_L^{\dagger})_{i_L' i}\right) \left(\sum_{i_L} (P_L)_{i i_L} \psi_{i_L n}\right) \nonumber \\
             & = \sum_{i} \tilde{\psi}^*_{i n'} \tilde{\psi}_{in} \,, \label{eq:m_explicit}
\end{align}
where in the last line we introduced the (un-physical) wave function $|\tilde{\psi}\rangle = (P_L \otimes \mathbb{I}) \ket{\psi}$.
Then, for $z \in \mathbb{C}^n$,
\begin{align*}
    z^{\dagger} M z & = \sum_{n' n} \sum_{i} \left(z^*_{n'} \tilde{\psi}^*_{i n'}\right) \left(\tilde{\psi}_{i n} z_n\right) \\
                    & = \sum_{i} \left(\sum_{n'} z^*_{n'} \tilde{\psi}^*_{i n'} \right) \left(\sum_{n} \tilde{\psi}_{i n} z_n\right) \\
                    & = \sum_{i} \left| \sum_{n} \tilde{\psi}_{i n} z_n \right|^2 \geq 0 \,,
\end{align*}
and therefore $M$ is positive semi-definite. \\
This result can be generalized to arbitrary quantum tensor, also including classical indices.
The measurement matrix $M$ can be written as 
\begin{equation}
    M^{n' n} = \sum_{\bm{j}} \left(\sum_{\bm{i}'} (\psi^{n'}_{\bm{i}'})^* (P_{\bm{j} \bm{i}'})^* \right) \left(\sum_{\bm{i}} (P)_{\bm{j} \bm{i}} \psi^n_{\bm{i}}\right) \,,
\end{equation}
where $\bm{i} = \{ i_1, i_2, \ldots, i_k \}$ are the indices of the quantum tensor which are contracted over, $\bm{j} = \{ j_1, j_2, \ldots, j_k \}$ are the indices to perform the contraction between the $\{P_l\}$ matrices, and $\{n', n\}$ is either a quantum or classical index that is left uncontracted.
Since this expression has the same structure as Eq.~\eqref{eq:m_explicit}, this also leads to the same conclusion that $M$ is positive semi-definite.

\section{Local diagonalization of observables}\label{app:local_diag_of_obs}
In the following, we illustrate how the local diagonalization of an observable facilitates calculating its expectation value with respect to a quantum tensor. 
We consider a quantum tensor with $k$ quantum indices $i_l$ of bond dimensions $\chi_l$, where $l \in \{1, 2, \ldots, k\}$ and $i_l \in \{1, 2, \ldots \chi_l\}$. 
We assume that the bond dimension is an integer power of 2, and therefore the number of qubits representing the quantum links is $n_{\text{qubits}}^{(l)} = \log_2 \chi_l$.
The observable can be written as a tensor product of local observables defined on each of the links
\begin{equation}
    \hat{O} = \hat{O}_1 \otimes \hat{O}_2 \otimes \cdots \otimes \hat{O}_k \,.
\end{equation}
In order to measure the expectation value of this observable with respect to the quantum tensor, each local observable $\hat{O}_l$ is decomposed as a linear combination of Pauli strings
\begin{equation}
    \hat{O}_l = \sum\limits_{j=1}^{N_l} c^{(l)}_j \hat{P}_j^{(l)} \,,
\end{equation}
where $\hat{P}^{(l)}_j \in \{\mathbb{I}, X, Y, Z\}^{\otimes n_{\text{qubits}}^{(l)}}$ is a tensor product of Pauli operators.
In the worst case, the number of terms in the sum is $N_l = 4^{n^{(l)}_{\text{qubits}}} = \chi_l^2$.
Therefore, the upper bound on the total number of terms in the decomposition of $\hat{O}$ is 
\begin{equation}
    N = \prod_{l=1}^k N_l = \prod_{l=1}^k \chi_l^2 \,.
\end{equation}
If we assume that all links have the same bond dimension $\chi$, then $N = \chi^{2k}$.
The number of terms in the decomposition of $\hat{O}$ therefore grows exponentially when increasing the number of legs of the quantum tensors.
Since the decomposition involves all elements of the Pauli basis, the number of measurements needed to measure the expectation value also grows exponentially with the number of legs of the quantum tensor, and becomes prohibitive even when applying advanced measurement strategies like, e.g., classical shadows~\cite{huang2020predicting} or adaptive POVMs~\cite{garcia2021learning}.

One possible strategy to reduce the number of measurements is to locally apply a unitary transformation to each observable
\begin{equation}
    \hat{O}_l = U^\dagger_l \hat{D}_l U_l \,.
\end{equation}
The classical cost of performing this decomposition is $O(\chi_l^3)$.
The unitaries $U_l$ can be appended to the quantum circuit representing the quantum tensor
\begin{equation}
    | \tilde{\psi} \rangle = (U_1 \otimes U_2 \otimes \cdots \otimes U_k) U(\theta) \ket{0} \,,
\end{equation}
where $U(\theta)$ is the parameterized unitary of the quantum tensor.
The decomposition of the diagonal matrix $\hat{D}_l$, can now be expressed in terms of Pauli strings $\hat{P}^{(l)}_j \in \{\mathbb{I}, Z\}^{\otimes n_{\text{qubits}}^{(l)}}$, which leads to at most $N'_i = 2^{n_{\text{qubits}^{(l)}}} = \chi_l$ terms in the decomposition.
The total number of terms in the decomposition of $\hat{D} = \hat{D}_1 \otimes \hat{D}_2 \otimes \cdots \otimes \hat{D}_k$ is $N' = \prod_{l=1}^k \chi_l$, which still grows exponentially when increasing the number of legs of the quantum tensor.
However, since all terms in the decomposition commute, only a single measurement is required to evaluate the expectation value $\langle \tilde{\psi} | \hat{D} | \tilde{\psi} \rangle$.
It is therefore possible to exponentially reduce the number of measurements at the cost of implementing the layer of unitaries 
\begin{equation}
    U_1 \otimes U_2 \otimes \cdots \otimes U_k \,.
\end{equation}
To implement each $n^{(l)}_{\text{qubits}}$-qubit unitary $U_l$ as a quantum circuit, the depth of the circuits scales in the worst case as $O(e^{n^{(l)}_{\text{qubits}}})$~\cite{krol2022efficient}.
However, in terms of the bond dimension, the scaling is $O(\chi_l)$. 
Since all the unitaries can be prepared in parallel, the quantum circuit that implements the layer of unitaries has a depth that scales with $O(\chi_\text{max})$, where $\chi_\text{max} = \max\limits_l \chi_l$.

The classical cost of the local diagonalization is therefore $O(k \chi^3)$ (assuming the same bond dimension $\chi$ for all $k$ legs of the quantum tensors). The additional quantum cost is $O(\chi)$ due to implementing the diagonalization unitaries as quantum circuits.

\section{Contraction order in hTNs}\label{app:contraction_order}

\begin{figure}
    \centering
    \includegraphics[width=\columnwidth]{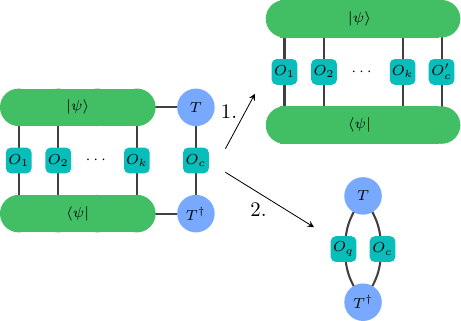}
    \caption{
        Comparison of contraction paths when calculating the expectation value of an observable with a hTN consisting of a one quantum and one classical tensor.
        Contraction path 1. is first contracting the classical tensor with the observable, and then the contraction involving the quantum tensor.
        Contraction path 2. is first contracting the quantum tensor with the observable, and then the contraction involving the classical tensor.
    }
    \label{fig:contraction_order}
\end{figure}

In this section, we illustrate how the contraction order affects the scaling of performing operations on a hTN. 
The following considerations generalize the statements in Ref.~\cite{yuan2021quantum}.

We consider a minimal hTN with two tensors, a rank-($k+1$) quantum tensor and a rank-2 classical tensor connected via a link (see Fig.~\ref{fig:contraction_order} for illustration).
The goal is to calculate the expectation value of an observable $O = O_1 \otimes O_2 \otimes \cdots \otimes O_k \otimes O_c$, where the local operators $O_i$ with $i= 1, \ldots, k$ contract with the legs of the quantum tensor, and $O_c$ contracts with the leg of the classical tensor.
We have two different contraction orders for the calculation of this expectation value.

Option 1: first contract the observable $O_c$ with the classical tensor $T$ to obtain a new observable $(O_c')^{i' i} = \sum_{j' j} (T^\dagger)^{i' j'} O_c^{j' j} T^{ij}$.
This observable is then used to measure the expectation value
\begin{equation}
    \bra{\psi} O_1 \otimes O_2 \otimes \cdots \otimes O_k \otimes O_c' \ket{\psi} \,.
\end{equation}
Applying the local diagonalization of observables introduced in the previous section, the number of measurements to estimate this expectation value is
\begin{equation}
    N_{\text{meas}} = M \cdot O(1) \,, 
\end{equation}
where $M$ is the number of samples used to measure single expectation values. The $O(1)$ scaling follows from the observation that all terms in the decomposition of the observable commute after performing the local diagonalization.

Option 2: first contract the quantum tensor with the corresponding part of the observable to obtain the operator
\begin{equation}
    (O_q')^{i' i} = \bra{\psi} O_1 \otimes O_2 \otimes \cdots \otimes O_k \otimes (\ketbra{i'}{i}) \ket{\psi} \,.
\end{equation}
For the evaluation of this expectation value, the number of measurements is
\begin{equation}
    N_{\text{meas}} = M N_{\text{op},q} \cdot O(1) \,,
\end{equation}
where $N_{\text{op},q} = 3^q$ is the number of expectation values required for the reconstruction of $O_q'$ and the $O(1)$ term follows from the same reasoning as above.
The observable $O_q'$ is then used to perform the contraction involving the classical tensor.

Therefore, option 1 has a much more favourable scaling and, therefore, should be preferred.

\end{document}